\documentclass[prb,twocolumn,showpacs]{revtex4}

\usepackage{graphicx,epsf}% Include figure files
\usepackage{bm}% bold math
 
\begin{document}

\title{Phase diagram of the anisotropic multichannel Kondo
       Hamiltonian revisited}
\smallskip

\author{Avraham Schiller$^{1}$ and Lorenzo De Leo$^2$}

\affiliation{$^1$Racah Institute of Physics, The Hebrew
                  University, Jerusalem 91904, Israel\\
             $^2$Center for Materials Theory, Serin
                  Physics Laboratory, Rutgers University,
                  136 Frelinghuysen Road,
                  Piscataway, NJ 08854-8019 USA}

\begin{abstract}
The phase diagram of the multichannel Kondo Hamiltonian with
an XXZ spin-exchange anisotropy is revisited, revealing a far
richer fixed-point structure than previously appreciated. For
a spin-$\frac{1}{2}$ impurity and $k > 2$ conduction-electron
channels, a second ferromagnetic-like domain is found deep
in the antiferromagnetic regime. The new domain extends
above a (typically large) critical longitudinal coupling
$J_z^{\ast} > 0$, and is separated from the antiferromagnetic
domain by a second Kosterliz-Thouless line. A similar line
of stable ferromagnetic-like fixed points with a residual
isospin-$\frac{1}{2}$ local moment is shown to exist for
large $J_z \gg |J_{\perp}| > 0$ and arbitrary $k$ and $s$
obeying $|k - 2s| > 1$. Here $J_z$ is the longitudinal
spin-exchange coupling, $J_{\perp}$ is the transverse
coupling, and $s$ is the impurity spin. Near the
free-impurity fixed-point, spin-exchange anisotropy is a
relevant perturbation for $s > 1/2$ and arbitrary $k$.
Depending on the sign of $J_z^2 - J_{\perp}^2$ and the parity
of $2s$, the system flows either to a conventional Fermi
liquid with no residual degeneracy, or to a $k$-channel,
spin-$\frac{1}{2}$ Kondo effect, or to a line of
ferromagnetic-like fixed points with a residual
isospin-$\frac{1}{2}$ local moment. These results
are obtained through a combination of perturbative
renormalization-group techniques, Abelian bosonization,
a strong-coupling expansion in $1/J_z$, and explicit
numerical renormalization-group calculations.
\end{abstract}

%% 75.20.Hr : Local moment in compounds and alloys; Kondo effect,
%%            valence fluctuations, heavy fermions
%% 72.15.Qm : Scattering mechanisms and Kondo effect.

\maketitle

\section{Introduction and overview of results}
%% \section{Introduction}

For over the last forty years, the Kondo effect has
occupied a central place in condensed matter physics.
While earlier studies of the effect have focused on
its single-channel version realized in dilute magnetic
alloys and valence-fluctuating systems, later attention
has largely shifted to its more exotic multichannel
variants where deviations from conventional Fermi-liquid
behavior can be found. The overscreened Kondo effect
is a paradigmatic example for quantum criticality in
quantum-impurity systems. Besides the possible
relevance to certain heavy fermion
alloys,~\cite{Cox87,Seaman_etal_91} ballistic metal point
contacts,~\cite{RB92,RLvDB94} scattering off two-level
tunneling systems,~\cite{Zawadowski80,VZ83,ThAsSe} and the
charging of small metallic grains,~\cite{Matveev91,Ashoori99}
the overscreened Kondo effect is one of the rare examples
of interaction-driven
non-Fermi-liquid behavior that is well understood and
well characterized theoretically.~\cite{CZ98} The
underscreened Kondo effect, which might be realized
in ultrasmall quantum dots with an even number of
electrons,~\cite{PC05} is a prototype for yet
another form of unconventional behavior --- that of
a singular Fermi liquid.~\cite{Mehta-etal05}

These vastly different ground states, as well as that of
an ordinary Fermi liquid, can all be described within the
single framework of the multichannel Kondo Hamiltonian,
which is among the simplest yet richest models for
strong electronic correlations in condensed-matter
physics. The multichannel Kondo Hamiltonian of
Eq.~(\ref{XXZ-Kondo}) describes the spin-exchange
interaction of a spin-$s$ local moment with $k$
identical, independent bands of spin-$\frac{1}{2}$
electrons. For antiferromagnetic exchange, the low-energy
physics features a subtle interplay between $k$ and $s$,
which could be summarized as follows.~\cite{CZ98}
For $k = 2s$, the impurity spin is exactly screened. A
spin singlet progressively forms below a characteristic
Kondo temperature $T_K$, leading to the formation of a
local Fermi liquid. For $k > 2s$, the impurity spin is
overscreened by the $k$ conduction-electron channels.
The system flows to an intermediate-coupling,
non-Fermi-liquid fixed point, characterized by anomalous
thermodynamic and dynamical properties. The elementary
excitations are collective in nature, in contrast to the
concept of a Fermi liquid. For $k < 2s$, the impurity
spin is underscreened. The low-energy physics comprises of
quasiparticle excitations plus a residual moment of size
$s' = s - k/2$. However, it differs from a conventional 
Fermi liquid in the singular energy dependence
of the scattering phase shift and divergence of the
quasiparticle density of
states.~\cite{Mehta-etal05,CP03,KHM05}
Such behavior was recently termed a singular Fermi
liquid.~\cite{Mehta-etal05} Similar qualitative behavior
is found for ferromagnetic exchange with arbitrary $k$
and $s$, except that the residual moment has the full
spin $s$.

Inherent to some of the leading scenarios for the
realization of the multichannel Kondo
model~\cite{Zawadowski80,VZ83,Matveev91} is a large
spin-exchange anisotropy. An XXZ anisotropy is well known
to be irrelevant both in the single-channel ($k = 1$,
$s = 1/2$) and the two-channel ($k = 2$, $s = 1/2$) cases.
The accepted phase diagram for these two models consists
of an antiferromagnetic and a ferromagnetic domain,
separated by a Kosterliz-Thouless line that traces
$J_z = -|J_{\perp}|$ at weak coupling. Here $J_z$ and
$J_{\perp}$ are the longitudinal and transverse exchange
couplings, respectively. As long as one lies within the
confines of the antiferromagnetic domain, the system
flows to the isotropic spin-exchange fixed point
regardless of how large the exchange anisotropy is.

Far less explored is the role of spin-exchange anisotropy
for either $s > 1/2$ or $k >2$. The stability of the
overscreened fixed point against weak spin-exchange
anisotropy has been analyzed in Ref.~\onlinecite{ALPC92}
using conformal field theory. For $k > 1$ and either
$s = 1/2$ or $s = (k - 1)/2$, the non-Fermi-liquid fixed
point was found to be stable against a weak spin-exchange
anisotropy. This has led to the perception that
spin-exchange anisotropy is irrelevant for these values
of $k$ and $s$. In contrast, spin-exchange anisotropy is
a relevant perturbation at the overscreened fixed point
for all other values of $1/2 < s < (k - 1)/2$
(assuming $k > 4$; for $k=4$ it is a marginal
perturbation),~\cite{ALPC92} though
the nature of the anisotropic fixed points was never
explored. Likewise unexplored is the effect of spin-exchange
anisotropy on the underscreened fixed point for $s > k/2$.

%% \subsection{Overview of results}

In this paper, we revisit the phase diagram of the
multichannel Kondo model with an XXZ spin-exchange
anisotropy. We find a far more complex picture than
previously appreciated, including new coupling regimes
where an XXZ anisotropy substantially alters the
low-energy physics. Our main findings are as follows.
\begin{itemize}
\item
    For $s = 1/2$ and $k >2$, a second ferromagnetic-like
    domain is found deep in the antiferromagnetic regime.
    The new domain extends above a (typically large)
    critical longitudinal coupling $J_z^{\ast} > 0$, and
    is separated from the conventional antiferromagnetic
    (non-Fermi-liquid) domain by a second Kosterliz-Thouless
    line.
\item
     For spin $s > 1/2$ and arbitrary $k$, spin-exchange
     anisotropy is relevant near the free-impurity fixed
     point. For sufficiently small
     $0 < |\rho J_{\perp}| < |\rho J_z| \ll 1/ks$
     ($\rho$ is the conduction-electron density of states
     per lattice site), the system flows to a line of stable
     ferromagnetic-like fixed points with a residual
     isospin-$\frac{1}{2}$ moment. For sufficiently small
     $|\rho J_z| < |\rho J_{\perp}| \ll 1/ks$,
     the flow is either to a conventional Fermi liquid
     with no residual degeneracy (for half-integer $s$),
     or to a $k$-channel Kondo effect with an effective
     spin-$\frac{1}{2}$ local moment (for integer $s$). Here
     by sufficiently small $\rho J_z$ and $\rho J_{\perp}$
     we mean the limit $J_z, J_{\perp} \to 0$ with any
     fixed ratio $r = J_z/J_{\perp} \neq \pm 1$. The closer
     $|r|$ is to one, the smaller the couplings must be for
     these results to apply.
\item
      For $|k - 2s| > 1$ and a sufficiently large
      $J_z \gg |J_{\perp}| > 0$, the system flows to a line
      of stable ferromagnetic-like fixed points with a
      residual isospin-$\frac{1}{2}$ moment. Only for
      $k = 2s$ and $k = 2s + 1$ are the exactly screened
      and overscreened fixed points stable against such a
      large anisotropy, which is marginal for $k = 2s -1$.
\end{itemize}

\begin{widetext}

\begin{figure}
\centerline{
\includegraphics[width=170mm]{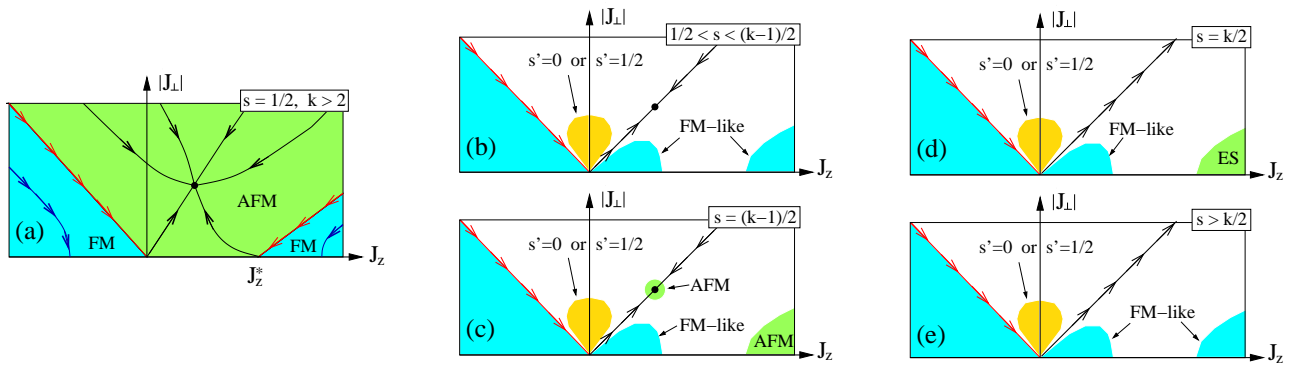}
}\vspace{-5pt}
\caption{(Color online) Phase diagram of the multichannel
         Kondo model for the different categories of $s$
         and $k$.
         (a) $s = 1/2$ and $k > 2$. In addition to the
         well-established ferromagnetic domain for
         $J_z \leq -|J_{\perp}|$, a second
         ferromagnetic-like domain extends to the right
         and below the red line stretching from
         $(J_z, J_{\perp}) = (J_z^{\ast}, 0)$. Within
         bosonization, $J_z^{\ast}$ is given by
         Eq.~(\ref{J-ast}).
         (b) $1/2 < s < (k-1)/2$. In addition to the
         conventional ferromagnetic domain for
         $J_z \leq -|J_{\perp}|$, the system flows to a
         line of stable ferromagnetic-like fixed points
         with a residual isospin-$\frac{1}{2}$ local moment
         in two opposite limits: (i) for sufficiently small
         $J_z > |J_{\perp}|$, (ii) for sufficiently large
         $J_z \gg |J_{\perp}|, D$ ($D$ being the
         conduction-electron bandwidth).
         For weak $|J_{\perp}| > |J_z|$, the system flows
         either to a Fermi-liquid fixed point with a frozen
         impurity (for integer $s$), or to an overscreened
         $k$-channel, spin-$\frac{1}{2}$ Kondo effect (for
         half-integer $s$). The above weak-coupling behavior
         is generic to $s > 1/2$. It extends to all cases
         described below, irrespective of $k$.
         (c) $s = (k-1)/2$. In contrast to (b), the
         isotropic overscreened fixed point is stable
         against a weak spin-exchange anisotropy
         (i.e., in the vicinity of the overscreened
         fixed point), as well as against a sufficiently
         large $J_z \gg |J_{\perp}|, D$.
         (d) $s = k/2$. The exactly screened fixed point,
         corresponding to strong coupling, is stable
         against a sufficiently large anisotropy,
         $J_z \gg |J_{\perp}|, D$. However, it is unstable
         at weak coupling, as described above.
         (e) $s > k/2$. All underscreened cases flow to a
         line of stable ferromagnetic-like fixed points
         with a residual isospin-$\frac{1}{2}$ local
         moment both for sufficiently large
         $J_z \gg |J_{\perp}|$ and for sufficiently small
         $J_z > |J_{\perp}|$.}
\label{fig:fig1}
\end{figure}

\end{widetext}

A compilation of these results is presented in
Fig.~\ref{fig:fig1}, in the form of phase diagrams for the
different categories of $s$ and $k$. The phase diagrams
for $s > 1/2$ are still incomplete. There remain extended
coupling regimes where the low-energy physics is yet to be
determined.

%% \subsection{Plan of the paper}

To obtain these results, we employ a combination of
perturbative renormalization-group (RG) techniques, Abelian
bosonization, a strong-coupling expansion in $1/J_z$, and
explicit numerical renormalization-group~\cite{Wilson75}
calculations. Using perturbative RG we first analyze in
Sec.~\ref{sec:Perturbative-RG} the limit of weak coupling.
In addition to the standard RG equations for the
dimensionless couplings $\rho J_z$ and $\rho J_{\perp}$,
a new Hamiltonian term proportional to $S_z^2$ is generated
for $s > 1/2$ and $J_z \neq J_{\perp}$. Here $S_z$ is the
$z$ component of the impurity spin. Depending on the sign
of $J_z^2 - J_{\perp}^2$, the new Hamiltonian term
favors either the maximally polarized impurity states
($S_z = \pm s$) or the minimally polarized ones.
This leads to a qualitative distinction between
$J_z > |J_{\perp}|$ and $|J_{\perp}| > |J_z|$, and to
the different low-temperature behaviors described above.

Proceeding with Abelian bosonization, we next show in
Sec.~\ref{sec:mapping} that the anisotropic multichannel
Kondo Hamiltonian with $s = 1/2$ and $k > 1$ possesses an
exact mapping between the two sets of coupling constants
$(J_z, J_{\perp})$ and $(J'_z, J_{\perp})$, where
\begin{equation}
\arctan \left (\frac{\pi \rho J'_z}{4} \right ) =
      \frac{\pi}{k}
      - \arctan \left (\frac{\pi \rho J_z}{4} \right ) .
\label{mapping}
\end{equation}
The above mapping is restricted to values of $J_z$ where
the right-hand side of Eq.~(\ref{mapping}) does not exceed
$\pi/2$. For $k = 2$, this condition limits the validity
of the mapping to $J_z \geq 0$, in which case
Eq.~(\ref{mapping}) simplifies to
\begin{equation}
\rho J'_z = \left (\frac{4}{\pi} \right )^2
            \frac{1}{\rho J_z} .
\label{2chK}
\end{equation}
For $k > 2$, the mapping extends to negative values of $J_z$,
relating $J_z > J_z^{\ast} = (4/\pi \rho) \tan (\pi/k)$
to $J_{\rm min} \leq J'_z < 0$ and vice versa [see
Eq.~(\ref{J_min}) for definition of $J_{\rm min}$].
Thus, in contrast to common perception, the multichannel
Kondo Hamiltonian with $s = 1/2$ and $k > 2$ possesses
a line of stable ferromagnetic-like fixed points for
$J_z > J_z^{\ast}$. Note, however, that $J_z^{\ast}$
exceeds the bandwidth for intermediate values of $k$,
and is pushed to weak coupling for $k \gg 1$.

Interestingly, the mapping of Eq.~(\ref{mapping}) is
ingrained in the Anderson-Yuval approach to the
multichannel Kondo problem, devised in
Refs.~\onlinecite{FGN95} and \onlinecite{Ye96}.
In fact, it was already recognized by Fabrizio
{\em et al.} for $k = 2$,~\cite{FGN95} but was never
appreciated to our knowledge for $k > 2$. Here we
provide an explicit operator mapping between the two
sets of model parameters, for arbitrary $k > 1$.

Since the critical coupling $J_z^{\ast}$ predicted by
Eq.~(\ref{mapping}) is typically larger than the bandwidth,
one might question the applicability of bosonization and
its unbound linear dispersion. Could it be that
$J_z^{\ast} \to \infty$ when the conduction electrons
are placed on a lattice? To eliminate this concern, a
strong-coupling expansion in $1/J_z$ is carried out in
Secs.~\ref{sec:SC-s=1/2} and \ref{sec:SC-general-s},
first for $s = 1/2$ and then for arbitrary $s$. The
strong-coupling expansion not only
confirms the existence of the new line of stable
ferromagnetic-like fixed points for $s = 1/2$ and
$k > 2$, but further extends this result to arbitrary
$s$ and $k$ obeying $|k - 2s| > 1$. For $k = 2s$ and
$k = 2s + 1$, it establishes the irrelevance of such
a large anisotropy.

As an explicit demonstration of these ideas,
the phase diagram of the $s = 1/2$, $k = 3$ model is
studied in Sec.~\ref{sec:NRG} using Wilson's numerical
renormalization-group (NRG) method.~\cite{Wilson75}
The NRG results confirm the phase diagram inferred from
Eq.~(\ref{mapping}), including the order of magnitude
of the critical coupling $J_z^{\ast}$. An extension
of the mapping of Eq.~(\ref{mapping}) to a spin-one
impurity is presented in turn in Appendix~\ref{app:s=1}.
We conclude in Sec.~\ref{sec:discussion} with a
discussion and summary of our results.

\section{Weak coupling}
\label{sec:Perturbative-RG}

We begin our discussion with the limit of weak coupling,
which is treated using perturbative RG. In the standard
notation, the multichannel Kondo Hamiltonian reads
\begin{eqnarray}
{\cal H} &=& \sum_{n=1}^{k}
             \sum_{\sigma = \uparrow, \downarrow}
             \sum_{\vec{k}} \epsilon_{\vec{k}}
               c^{\dagger}_{\vec{k} n \sigma}
               c_{\vec{k} n \sigma}
\nonumber \\
&+&
           \frac{J_z}{2N} \sum_{n=1}^{k}
           \sum_{\vec{k}, \vec{k}'}
           \left [
                   c^{\dagger}_{\vec{k} n \uparrow}
                   c_{\vec{k}' n \uparrow}
                 - c^{\dagger}_{\vec{k} n \downarrow}
                   c_{\vec{k}' n \downarrow}
           \right ] S_{z}
\nonumber \\
&+&
           \frac{J_{\perp}}{2N}
           \sum_{n=1}^{k} \sum_{\vec{k}, \vec{k}'}
           \left [
                   c^{\dagger}_{\vec{k} n \uparrow}
                   c_{\vec{k}' n \downarrow} S^{-}
                   + c^{\dagger}_{\vec{k} n \downarrow}
                   c_{\vec{k}' n \uparrow} S^{+}
           \right ]\!.
\label{XXZ-Kondo}
\end{eqnarray}
Here, $c^{\dagger}_{\vec{k} n \sigma}$ creates an electron
with wave number $\vec{k}$ and spin projection $\sigma$
in the $n$th conduction-electron channel; $\vec{S}$ is
a spin-$s$ operator; $J_z$ and $J_{\perp}$ are the
longitudinal and transverse Kondo couplings, respectively;
and $N$ is the number of lattice sites.

As a first step toward devising a perturbative RG
treatment of the Hamiltonian of Eq.~(\ref{XXZ-Kondo}), we
convert the model to dimensionless form. To this end,
we introduce the fermion operators
\begin{equation}
a^{\dagger}_{\varepsilon n \sigma} =
                 \frac{1}{\sqrt{D \rho(E) N}}
                      \sum_{\vec{k}}
                      \delta(\varepsilon-\epsilon_{\vec{k}}/D)
                      c^{\dagger}_{\vec{k} n \sigma} ,
\end{equation}
which represent the $2k$ conduction-electron modes
that couple to the impurity within the energy shell
$E = \varepsilon D$. Here $D$ is the conduction-electron
bandwidth and $\rho(E)$ is the conduction-electron density
of states per lattice site:
\begin{equation}
\rho(E) = \frac{1}{N} \sum_{\vec{k}}
                 \delta(E - \epsilon_{\vec{k}}) .
\end{equation}
The dimensionless operators $a^{\dagger}_{\varepsilon n \sigma}$
have been normalized to obey canonical anticommutation
relations:
\begin{equation}
\left \{
         a_{\varepsilon n \sigma},
         a^{\dagger}_{\varepsilon' n' \sigma'}
\right \} =
         \delta_{n n'} \delta_{\sigma \sigma'}
         \delta(\varepsilon - \varepsilon') .
\end{equation}
Assuming a box density of states
$\rho(E) = \rho \theta(D - |E|)$ and omitting all modes
that decouple from the impurity, Eq.~(\ref{XXZ-Kondo})
is recast in the form ${\cal H} = D \tilde{\cal H}$,
where $\tilde{\cal H}$ is the dimensionless Hamiltonian
\begin{eqnarray}
\tilde{\cal H} &=& \sum_{n = 1}^{k}
                   \sum_{\sigma = \uparrow, \downarrow}
                   \int_{-1}^{1} \varepsilon
                          a^{\dagger}_{\varepsilon n \sigma}
                          a_{\varepsilon n \sigma} d\varepsilon
\nonumber \\
&+&
                     \frac{\tilde{J}_z}{2}
                     \sum_{n=1}^{k}
                     \int_{-1}^{1}\!d\varepsilon
                     \int_{-1}^{1}\!d\varepsilon'
                     \left [
                          a^{\dagger}_{\varepsilon n \uparrow}
                              a_{\varepsilon n \uparrow}
                        - a^{\dagger}_{\varepsilon n \downarrow}
                              a_{\varepsilon n \downarrow}
                     \right ] S_z
\nonumber \\
&+& 
                   \frac{\tilde{J}_{\perp}}{2}
                   \sum_{n=1}^{k} 
                   \int_{-1}^{1}\!d\varepsilon
                   \int_{-1}^{1}\!d\varepsilon'
                   \left [
                        a^{\dagger}_{\varepsilon n \uparrow}
                        a_{\varepsilon n \downarrow} S^{-}
                        + {\rm H.c.}
                   \right ] .
\label{dimensionless-H}
\end{eqnarray}
Here $\tilde{J}_{z} = \rho J_z$ and $\tilde{J}_{\perp} =
\rho J_{\perp}$ are the dimensionless Kondo couplings.

Focusing on $|\tilde{J}_z|, |\tilde{J}_{\perp}| \ll
1/ks$, we treat the Hamiltonian of
Eq.~(\ref{dimensionless-H}) using perturbative RG. The
RG transformation consists of the following three steps.
Suppose that the bandwidth has already been lowered from
its initial value $D$ to some value $D' = D e^{-l}$
($l > 0$). Further lowering the bandwidth to $D'' =
D'(1 - \delta l)$ requires the elimination of all
$a_{\varepsilon n \sigma}$ degrees of freedom in the
interval $1 - \delta l < |\varepsilon| \leq 1$. This
goal is accomplished using Anderson's poor-man's
scaling.~\cite{Anderson70} At the conclusion of this
step, all integrations in Eq.~(\ref{dimensionless-H})
have been reduced to the range
$-(1 - \delta l) \leq x \leq 1 - \delta l$.
The RG transformation is completed by (i) rescaling
$\tilde{\cal H} \to \tilde{\cal H}/(1 - \delta l)$ to
account for the reduced bandwidth, and (ii) restoring the
original integration ranges in Eq.~(\ref{dimensionless-H})
by converting to
\begin{equation}
a_{\varepsilon n \sigma} \to
     \tilde{a}_{\varepsilon n \sigma} = (1-\delta l)^{1/2}
     a_{(1 - \delta l)\varepsilon n \sigma} .
\end{equation}
This latter step allows us to write
\begin{equation}
\int_{-(1 - \delta l)}^{1 - \delta l}
     a_{\varepsilon n \sigma} d\varepsilon =
(1 - \delta l)^{1/2} \int_{-1}^{1}
     \tilde{a}_{\varepsilon n \sigma} d\varepsilon .
\end{equation}
The dimensionless Hamiltonian is recast in this manner
in a self-similar form, but with renormalized couplings.
These obey a set of coupled differential equations,
specified below.

For $s = 1/2$ we recover the familiar RG equations
\begin{eqnarray}
\frac{d \tilde{J}_z}{dl} &=& \tilde{J}_{\perp}^2 ,
\label{RG-z} \\
\frac{d \tilde{J}_{\perp}}{dl} &=& \tilde{J}_{\perp}
                                   \tilde{J}_z ,
\label{RG-perp}
\end{eqnarray}
showing that spin-exchange anisotropy is irrelevant for
weak $\tilde{J}_z > -|\tilde{J}_{\perp}|$, irrespective
of the number of channels $k$. A different qualitative
picture emerges for spin $s$ larger than one-half. Here
a new Hamiltonian term
$\tilde{\Delta} S_z^2$ is generated within
$\tilde{\cal H}$. Starting from zero, $\tilde{\Delta}$
renormalizes according to
\begin{equation}
\frac{d \tilde{\Delta}}{d l} = \tilde{\Delta} -
        k \ln(2) \left [
                \tilde{J}_z^2 - \tilde{J}_{\perp}^2
        \right ] ,
\label{d-Delta-dl}
\end{equation}
which supplements Eqs.~(\ref{RG-z}) and (\ref{RG-perp})
for $\tilde{J}_z$ and $\tilde{J}_{\perp}$. Restricting
attention to $s > 1/2$, we now analyze the ramifications
of the new coupling $\tilde{\Delta}$ as a function of $s$,
$k$, and the bare Kondo couplings $J_z$ and $J_{\perp}$.

Since $\tilde{J}_z^2 - \tilde{J}_{\perp}^2$ is conserved
under the RG, a negative (positive) $\tilde{\Delta}$ is
generated if the bare Kondo couplings satisfy
$|J_z| > |J_{\perp}|$ ($|J_z| < |J_{\perp}|$). For a
given ratio $J_z/J_{\perp} \neq \pm 1$ and sufficiently
small $|J_{\perp}|$, the coupling $|\tilde{\Delta}|$
approaches unity well before any Kondo temperature can
be reached. This has the effect of freezing all but the
lowest-lying spin states. For $|J_z| > |J_{\perp}|$
(negative $\tilde{\Delta}$), only the maximally polarized
states $S_z = \pm s$ are thus left. For $|J_z| < |J_{\perp}|$
(positive $\tilde{\Delta}$), the picture depends on $s$:
for half-integer $s$, the two degenerate states
$S_z = \pm 1/2$ are selected; for integer $s$, only the
state $S_z = 0$ remains.

Depending on which case is realized, a different
effective low-energy Hamiltonian emerges. Consider
first the case $|J_z| > |J_{\perp}|$ (negative
$\tilde{\Delta}$). Introducing the isospin operators
\begin{equation}
\tau_z = \frac{1}{2} |s\rangle \langle s| -
         \frac{1}{2} |-s\rangle \langle -s| ,
\label{tau-z-def}
\end{equation}
\begin{equation}
\tau^{\pm} = |\pm s\rangle \langle \mp s| ,
\label{tau-perp-def}
\end{equation}
the resulting low-energy Hamiltonian contains the
term
\begin{equation}
\tilde{\cal H}_{J_z} = s \tilde{J}_z
                \sum_{n=1}^{k}
                \int_{-1}^{1}\!d\varepsilon
                \int_{-1}^{1}\!d\varepsilon'
                \left [
                      a^{\dagger}_{\varepsilon n \uparrow}
                         a_{\varepsilon n \uparrow}
                    - a^{\dagger}_{\varepsilon n \downarrow}
                         a_{\varepsilon n \downarrow}
                \right ] \tau_z ,
\label{H_jz_1}
\end{equation}
which follows from projection of the spin-exchange
interaction of Eq.~(\ref{dimensionless-H}) onto the
$S_z = \pm s$ subspace. Here ${\tilde{J}}_z$ is the running
coupling constant at the scale where $|\tilde{\Delta}|$
approaches one. In contrast to Eq.~(\ref{H_jz_1}), terms
that flip the isospin $\vec{\tau}$ involve the creation
and annihilation of at least $2s$ conduction electrons.
This follows from the fact that a flip in $\vec{\tau}$
changes $S_z$ by $\pm 2s$. Since the Hamiltonian conserves
the total spin projection of the entire system (impurity
plus conduction electrons), such a process must be
accompanied by an opposite spin flip of $2s$ conduction
electrons in the vicinity of the impurity. As result,
isospin-flip terms have the scaling dimension
$\Delta_{\perp} \ge 2s \ge 2$ about the free-impurity
fixed point, rendering them irrelevant. The
resulting fixed-point structure corresponds then to a
line of ferromagnetic-like fixed points with a residual
isospin-$\frac{1}{2}$ local moment, characterized by
a finite $\tilde{J}_z$. Note that this picture is
insensitive to the sign of the Kondo couplings. It equally
applies to positive and negative $J_z$ and $J_{\perp}$.

Proceeding with the case $|J_z| < |J_{\perp}|$ (positive
$\tilde{\Delta}$), we first consider half-integer $s$.
Here the two states selected by $\tilde{\Delta}$ are
$S_z = \pm 1/2$. Similar to Eqs.~(\ref{tau-z-def}) and
(\ref{tau-perp-def}), we introduce the isospin operators
\begin{equation}
\tau_z = \frac{1}{2} |1/2\rangle \langle 1/2| -
         \frac{1}{2} |-1/2\rangle \langle -1/2| ,
\label{tau-z-def2}
\end{equation}
\begin{equation}
\tau^{\pm} = |\pm 1/2\rangle \langle \mp 1/2| ,
\label{tau-perp-def2}
\end{equation}
which now relate to the states $|S_z = \pm 1/2 \rangle$.
Projection of Eq.~(\ref{dimensionless-H}) onto the
$S_z = \pm 1/2$ subspace yields the following
effective spin-exchange interaction:
\begin{eqnarray}
\tilde{\cal H}_J &=&
            \frac{\tilde{J}_z}{2}
            \sum_{n=1}^{k}
            \int_{-1}^{1}\!d\varepsilon
            \int_{-1}^{1}\!d\varepsilon'
            \left [
                 a^{\dagger}_{\varepsilon n \uparrow}
                     a_{\varepsilon n \uparrow}
               - a^{\dagger}_{\varepsilon n \downarrow}
                     a_{\varepsilon n \downarrow}
            \right ] \tau_z
\nonumber \\
&+& 
            \gamma \frac{\tilde{J}_{\perp}}{2}
            \sum_{n=1}^{k} 
            \int_{-1}^{1}\!d\varepsilon\!
            \int_{-1}^{1}\!d\varepsilon'
            \left [
                 a^{\dagger}_{\varepsilon n \uparrow}
                 a_{\varepsilon n \downarrow} \tau^{-}\!
                 + {\rm H.c.}
            \right ]
\label{H_j_2}
\end{eqnarray}
with $\gamma = \sqrt{s(s+1) + 1/4} > 1$. Once again,
$\tilde{J}_z$ and $\tilde{J}_{\perp}$ in Eq.~(\ref{H_j_2})
are the running coupling constants at the scale where
$\tilde{\Delta}$ approaches one.

Equation~(\ref{H_j_2}) has the form of a $k$-channel
Kondo Hamiltonian with an effective isospin-$\frac{1}{2}$
local moment. Since $|\tilde{J}_z| < \gamma
|\tilde{J}_{\perp}| \ll 1$, one lies in the confines
of the antiferromagnetic domain. Hence, irrespective
of the original spin $s$, the system flows to the
overscreened fixed point of the $k$-channel Kondo
effect with spin $s' = 1/2$. (For $k = 1$, the flow is
to the strong-coupling fixed point of the conventional
one-channel Kondo effect). Excluding the case
$s = k/2 -1/2$ with antiferromagnetic exchange, the
resulting low-energy physics differs markedly from
that of the isotropic model, whether ferromagnetic
or antiferromagnetic. Most strikingly, the
underscreened fixed point for $J_z = J_{\perp} > 0$
and $s > k/2 > 1/2$ gives way to an overscreened fixed
point for any given ratio $-1 < J_z/J_{\perp} < 1$ and
sufficiently small $|J_{\perp}|$.

Of the different possible cases for $\tilde{\Delta}$
and $s$, the simplest picture is recovered for
$|J_z| < |J_{\perp}|$ and integer $s$. Here the
impurity spin is frozen in the $S_z = 0$ state,
loosing its dynamics. This results in a
conventional Fermi-liquid fixed point with neither
non-Fermi-liquid characteristics nor a residual
local-moment degeneracy.

\section{Exact mapping for $s = 1/2$}
\label{sec:mapping}

As is evident from the discussion above, there is a
qualitative difference between $s = 1/2$ and $s > 1/2$
concerning the effect of spin-exchange anisotropy for
weak coupling. In this section we focus on $s = 1/2$
and use Abelian bosonization to derive the mapping
of Eq.~(\ref{mapping}). Our starting point is the
continuum-limit representation of the multichannel
Kondo Hamiltonian
\begin{eqnarray}
{\cal H} &=& \sum_{n = 1}^{k}
             \sum_{\sigma = \uparrow, \downarrow}
             i \hbar v_F
             \int_{-\infty}^{\infty}
                  \psi^{\dagger}_{n \sigma}(x)
                  \partial_x \psi_{n \sigma}(x) dx
\label{H} \\
        &+& \sum_{n = 1}^{k}
                  \frac{J_{\perp} a}{2} \left [
                  \psi^{\dagger}_{n \downarrow}(0)
                       \psi_{n \uparrow}(0) S^{+}
                  + S^{-} \psi^{\dagger}_{n \uparrow}(0)
                       \psi_{n \downarrow}(0)
                  \right ]
\nonumber \\
        &+& \sum_{n = 1}^{k}
                  \frac{J_z a}{2} S_z\! \left [
                  \psi^{\dagger}_{n \uparrow}(0)
                       \psi_{n \uparrow}(0)\!
                  -\! \psi^{\dagger}_{n \downarrow}(0)
                         \psi_{n \downarrow}(0)
                  \right ]
        - g_i h S_z
\nonumber \\
        &-& \frac{g_e h}{2} \sum_{n = 1}^{k}
                  \int_{-\infty}^{\infty} \left [
                  \psi^{\dagger}_{n \uparrow}(x)
                       \psi_{n \uparrow}(x)
                  - \psi^{\dagger}_{n \downarrow}(x)
                         \psi_{n \downarrow}(x)
                  \right ] dx ,
\nonumber
\end{eqnarray}
written in terms of the left-moving one-dimensional fields
$\psi^{\dagger}_{n \sigma}(x)$ with $n = 1,\cdots,k$
and $\sigma = \uparrow, \downarrow$. Here, $\vec{S}$ is
a spin-$\frac{1}{2}$ operator; $a$ is a short-distance
cutoff corresponding to a lattice spacing; $x$ is a
fictitious coordinate conjugate to $k = \varepsilon \pi/a$;
and $h$ is an applied magnetic field. For the sake of
generality, we allow for different impurity and
conduction-electron Land\'e $g$-factors, $g_i$ and $g_e$,
respectively.

To treat the Hamiltonian of Eq.~(\ref{H}), we resort
to Abelian bosonization. According to standard
prescriptions,~\cite{Haldane81} $2k$ boson fields are
introduced --- one boson field $\Phi_{n \sigma}(x)$ for
each left-moving fermion field $\psi_{n \sigma}(x)$.
The fermion fields are written as
\begin{equation}
\psi_{n \sigma}(x) = P_{n \sigma}
             \frac{1}{\sqrt{2 \pi \alpha}}
             e^{-i\Phi_{n \sigma}(x)} ,
\label{psi-via-phi}
\end{equation}
where the $\Phi_{n \sigma}$ fields obey
\begin{equation}
\left [ \Phi_{n \sigma}(x),
        \Phi_{n' \sigma'}(y) \right] =
        -i \delta_{n n'} \delta_{\sigma \sigma'}
           \pi\, {\rm sign} (x-y) .
\end{equation}
The ultraviolet momentum cutoff $\alpha^{-1} = \pi/a$
is related to the conduction-electron bandwidth
$D$ and the density of states per lattice site
$\rho$ through $D = \hbar v_F/\alpha$ and $\rho =
1/(2 D) = \alpha/(2\hbar v_F)$, respectively. The
operators $P_{n \sigma}$ in Eq.~(\ref{psi-via-phi}) are
phase-factor operators, which come to ensure that the
different fermion species anticommute. Our explicit
choices for these operators are
\begin{equation}
P_{n \sigma} = e^{i \pi
                    \left [
                            N{n \uparrow} +
                            \sum_{j < n} \sum_{\sigma'}
                                          N_{j \sigma'}
                    \right ]} ,
\end{equation}
where $N_{j \sigma}$ is the number operator for electrons
in channel $j$ with spin projection $\sigma$.

In terms of the boson fields, the multichannel Kondo
Hamiltonian takes the form
\begin{eqnarray}
{\cal H} &=& \sum_{n = 1}^{k}
             \sum_{\sigma = \uparrow, \downarrow}
             \frac{\hbar v_F}{4 \pi}
             \int_{-\infty}^{\infty}
                  \left (
                        \nabla \Phi_{n \sigma}(x)
                  \right )^2 dx
\nonumber \\
&+& \sum_{n = 1}^{k} \frac{J_{\perp}}{2}
             \left \{
                  e^{i [\Phi_{n \uparrow}(0) -
                        \Phi_{n \downarrow}(0)]} S^{-}
                  + {\rm H.c.}
             \right \}
\nonumber \\
&+& \delta_z\frac{a}{\pi^2 \rho} \sum_{n = 1}^{k}
             \left [
                    \nabla \Phi_{n \uparrow}(0)
                    - \nabla \Phi_{n \downarrow}(0)
             \right ] S_z - g_i h S_z 
\nonumber \\
&-& \frac{g_e h}{4\pi} \sum_{n = 1}^{k}
                 \int_{-\infty}^{\infty}
                 \left [
                       \nabla \Phi_{n \uparrow}(x)
                       - \nabla \Phi_{n \downarrow}(x)
                 \right ] dx ,
\label{H-bosonized}
\end{eqnarray}
where
\begin{equation}
\delta_z = \arctan \left (\frac{\pi \rho J_z}{4} \right )
\end{equation}
is the parallel-spin phase shift induced by $J_z$ in the
absence of $J_{\perp}$. Note that $\delta_z$ is bounded in
magnitude by $\pi/2$, which stems from the cutoff scheme
used in bosonization. Although the bosonic Hamiltonian
of Eq.~(\ref{H-bosonized}) does support larger values of
$|\delta_z|$, this parameter must not exceed $\pi/2$ in
order for ${\cal H}$ to possess a fermionic counterpart
of the form of Eq.~(\ref{H}). We shall return to this
important point later on.

At this stage we manipulate the bosonic Hamiltonian of
Eq.~(\ref{H-bosonized}) through a sequence of steps. We
begin by converting to $2k$ new boson fields, consisting
of
\begin{equation}
\Phi_s(x) = \frac{1}{\sqrt{2k}} \sum_{n = 1}^{k}
            \left [ \Phi_{n \uparrow}(x)
                  - \Phi_{n \downarrow}(x)
            \right ]
\label{Phi_s-def}
\end{equation}
plus $2k - 1$ orthogonal fields: $\Phi_{\mu}(x)$ with
$\mu = 1,\cdots, 2k-1$. The orthogonal fields
$\Phi_{\mu}(x)$ are formally expressed as
\begin{equation}
\Phi_{\mu}(x) = \sum_{n = 1}^{k}
                \sum_{\sigma = \uparrow, \downarrow}
                   e^{\mu}_{n \sigma} \Phi_{n \sigma}(x) ,
\label{Phi_mu-def}
\end{equation}
where the real coefficients $e^{\mu}_{n \sigma}$ obey
\begin{equation}
\sum_{n = 1}^{k} \sum_{\sigma = \uparrow, \downarrow}
           e^{\mu}_{n \sigma} e^{\nu}_{n \sigma} = 
           \delta_{\mu \nu} ,
\end{equation}
\begin{equation}
\sum_{n = 1}^{k}
           \left [
                 e^{\mu}_{n \uparrow}
                 - e^{\mu}_{n \downarrow}
           \right ] = 0 .
\end{equation}
The precise form of the coefficients $e^{\mu}_{n \sigma}$
is of no practical importance to our discussion, and need
not concern us. Their choice is not unique. In terms of
the new fields, the combinations
$\Phi_{n \uparrow}(x) - \Phi_{n \downarrow}(x)$ take the form
\begin{equation}
\Phi_{n \uparrow}(x) - \Phi_{n \downarrow}(x) =
\sqrt{\frac{2}{k}} \Phi_s(x) + \varphi_n(x) ,
\end{equation}
where $\varphi_n(x)$ is some linear combination of the
fields $\Phi_{\mu}(x)$ with $\mu = 1,\cdots, 2k-1$.
Once again, the precise form of the $\varphi_n(x)$
combinations is of no real significance to our discussion.
We shall only rely on them being orthogonal to
$\Phi_{s}(x)$.

Using the new boson fields defined above, the Hamiltonian
of Eq.~(\ref{H-bosonized}) is converted to
\begin{eqnarray}
{\cal H} &=&
             \frac{\hbar v_F}{4 \pi}
             \int_{-\infty}^{\infty}
             \left [
                  \left (
                        \nabla \Phi_{s}
                  \right )^2
                  + \sum_{\mu = 1}^{2k-1}
                  \left (
                        \nabla \Phi_{\mu}
                  \right )^2
             \right ] dx
\nonumber \\
&+& \sum_{n = 1}^{k} \frac{J_{\perp}}{2}
             \left \{
                  e^{i\sqrt{\frac{2}{k}} \Phi_s(0)
                     + i\varphi_n(0)} S^{-}
                  + {\rm H.c.}
             \right \}
\nonumber \\
&+& \delta_z \frac{a}{\pi^2 \rho} \sqrt{2k}
                    \nabla \Phi_{s}(0) S_z
             - g_i h S_z 
\nonumber \\
&-& \frac{g_e h}{4\pi} \sqrt{2 k}
                 \int_{-\infty}^{\infty}
                       \nabla \Phi_{s}(x) dx .
\label{H-via-Phi_s}
\end{eqnarray}
Next the canonical transformation
${\cal H}' = U {\cal H} U^{\dagger}$ with
$U = \exp \left [i \sqrt{\frac{8}{k}} \Phi_s(0) S_z \right]$
is applied to obtain
\begin{eqnarray}
{\cal H}' &=& 
             \frac{\hbar v_F}{4 \pi}
             \int_{-\infty}^{\infty}
             \left [
                  \left (
                        \nabla \Phi_{s}
                  \right )^2
                  + \sum_{\mu = 1}^{2k-1}
                  \left (
                        \nabla \Phi_{\mu}
                  \right )^2
             \right ] dx
\nonumber \\
&+& \sum_{n = 1}^{k} \frac{J_{\perp}}{2}
             \left \{
                  e^{-i\sqrt{\frac{2}{k}} \Phi_s(0)
                     + i\varphi_n(0)} S^{-}
                  + {\rm H.c.}
             \right \}
\nonumber \\
&+& \left( \delta_z - \frac{\pi}{k} \right)
              \frac{a}{\pi^2 \rho} \sqrt{2k}
              \nabla \Phi_{s}(0) S_z
              - (g_i - 2g_e)  h S_z 
\nonumber \\
&-& \frac{g_e h}{4\pi} \sqrt{2 k}
                 \int_{-\infty}^{\infty}
                       \nabla \Phi_{s}(x) dx .
\label{H'}
\end{eqnarray}
Here we have omitted a constant term from ${\cal H}'$,
and made use of the identity $a/\pi \rho = 2\hbar v_F$
in writing the term $\nabla \Phi_s(0) S_z$.
Finally, the transformation $\Phi_s(x) \to -\Phi_s(x)$
is introduced, which yields
\begin{eqnarray}
{\cal H}' &=&
             \frac{\hbar v_F}{4 \pi}
             \int_{-\infty}^{\infty}
             \left [
                  \left (
                        \nabla \Phi_{s}
                  \right )^2
                  + \sum_{\mu = 1}^{2k-1}
                  \left (
                        \nabla \Phi_{\mu}
                  \right )^2
             \right ] dx
\nonumber \\
&+& \sum_{n = 1}^{k} \frac{J_{\perp}}{2}
             \left \{
                  e^{i\sqrt{\frac{2}{k}} \Phi_s(0)
                     + i\varphi_n(0)} S^{-}
                  + {\rm H.c.}
             \right \}
\nonumber \\
&+& \delta'_z \frac{a}{\pi^2 \rho} \sqrt{2k}
              \nabla \Phi_{s}(0) S_z
              + (2 g_e - g_i)  h S_z 
\nonumber \\
&+& \frac{g_e h}{4\pi} \sqrt{2 k}
                 \int_{-\infty}^{\infty}
                       \nabla \Phi_{s}(x) dx
\label{H-final}
\end{eqnarray}
with $\delta'_z = \pi/k - \delta_z$.

The Hamiltonian of Eq.~(\ref{H-final}) is identical to that
of Eq.~(\ref{H-via-Phi_s}), apart from a renormalization of
certain parameters: $\delta_z \to \delta'_z$, $h \to -h$, and
$g_i \to 2g_e - g_i$. As long as $|\delta'_z| \leq \pi/2$, one can
revert the series of steps leading to Eq.~(\ref{H-via-Phi_s}),
to recast the Hamiltonian of Eq.~(\ref{H-final}) in fermionic
form. The end result is just the original multichannel Kondo
Hamiltonian of Eq.~(\ref{H}) with the following renormalized
parameters:
\begin{equation}
J_z \to J'_z = \frac{4}{\pi \rho}
       \tan \left ( \frac{\pi}{k} - \delta_z \right) ,
\end{equation}
\begin{equation}
h \to -h ,
\end{equation}
\begin{equation}
g_i \to 2g_e - g_i .
\end{equation}

For zero magnetic field, this establishes the mapping of
Eq.~(\ref{mapping}), including the restriction to values of
$J_z$ where the right-hand side of Eq.~(\ref{mapping})
does not exceed $\pi/2$. The latter condition is just a
restatement of the requirement $|\delta'_z| \leq \pi/2$. We
now analyze in detail the ramifications of this restriction
as a function of the number of channels $k$.

For $k = 1$ (single-channel case), $\delta'_z$ exceeds
$\pi/2$ for all $-\infty < J_z < \infty$. Thus, the mapping
of Eq.~(\ref{mapping}) does not apply to the single-channel
Kondo effect, in accordance with known results. For
$k = 2$ (two-channel case), the required condition is
met for all $J_z \geq 0$, mapping weak to strong coupling
and vice versa [see Eq.~(\ref{2chK})]. Hence, the mapping
of Eq.~(\ref{mapping}) can be viewed as an anisotropic
variant~\cite{FGN95} of the weak-to-strong-coupling
duality of Nozi\'eres and Blandin.~\cite{NB80}

The most interesting case occurs for $k > 2$, when
Eq.~(\ref{mapping}) extends to all $J_z \geq J_{\rm min}$
with
\begin{equation}
J_{\rm min} = -\frac{4}{\pi \rho}
                 \tan \left (
                      \frac{\pi}{2} - \frac{\pi}{k}
                 \right ) < 0 .
\label{J_min}
\end{equation}
In particular, the range $J_z > J_z^{\ast}$ with
\begin{equation}
J_z^{\ast} = \frac{4}{\pi \rho} \tan
                \left (
                      \frac{\pi}{k}
                \right ) > 0
\label{J-ast}
\end{equation}
is mapped onto the negative-coupling regime $J_{\rm min}
\leq J'_z < 0$ and vice verse. Thus, the Kosterliz-Thouless
line separating the antiferromagnetic and ferromagnetic
domains is duplicated from $J_z = -|J_{\perp}|$ to
$J_z = J_z^{\ast} + C_k |J_{\perp}|$ with
\begin{equation}
C_k = \frac{1}
           {1 + \tan^2 \left( \pi/k \right )}
    = \frac{1}
           {1 + \left( \pi \rho J_z^{\ast}/4 \right )^2} .
\label{C_k}
\end{equation}
Here we have assumed
$\rho |J_z\!-\!J_z^{\ast}|, \rho|J_{\perp}| \ll 1$, in
order for the weak-coupling parameterization of the
Kosterliz-Thouless line to apply.

The resulting phase diagram for $k > 2$ is plotted in
Fig.~\ref{fig:fig1}(a). For $J_z > J_z^{\ast} + C_k |J_{\perp}|$,
the multichannel Kondo Hamiltonian flows to a line of
stable ferromagnetic-like fixed points, rendering the
non-Fermi-liquid fixed point of the model unstable against
a large enough anisotropy. This behavior contradicts
the common perception of spin-exchange anisotropy as
irrelevant for $s = 1/2$. Note that the same
phase diagram emerges from the renormalization-group
equations derived by Ye using the equivalent Anderson-Yuval
approach.~\cite{Ye96} However, the newly found line of
stable ferromagnetic-like fixed points has eluded Ye.

\section{Strong-coupling expansion for $s = 1/2$}
\label{sec:SC-s=1/2}

A potential concern with the above picture for $s = 1/2$
has to do with the validity of the bosonization approach
used. Since $J_z^{\ast}$ exceeds the bandwidth $D$ for
intermediate values of $k$, one might wonder to what
extent is bosonization (or the Anderson-Yuval approach
for that matter) justified for such strong coupling. In
fact, the very usage of the continuum-limit Hamiltonian
of Eq.~(\ref{H}) with its unbounded linear dispersion
can be called into question. Could it be that $J_z^{\ast}
\to \infty$ when the conduction electrons are placed on
a lattice?

Not withstanding the observation that $J_z^{\ast}$ for
$k \gg 1$ is pushed to weak coupling, to firmly establish
the phase diagram of Fig.~\ref{fig:fig1}(a) one must show
that the new line of stable ferromagnetic-like fixed
points extends to proper lattice models for the underlying
conduction bands. This is the objective of the following
analysis, which focuses on the limit of a large
longitudinal coupling, $J_z \gg D, |J_{\perp}|$.

As a generic lattice model for the conduction bands, we
consider a spin-$\frac{1}{2}$ impurity moment coupled to
the open end of a semi-infinite tight-binding chain with
$k$ identical conduction-electron species:
\begin{eqnarray}
{\cal H} &=&
         \sum_{j = 0}^{\infty}
         \sum_{n = 1}^{k}
         \sum_{\sigma = \uparrow, \downarrow}
         \left [
              \epsilon_j f^{\dagger}_{j n \sigma} f_{j n \sigma}
         \right .
\nonumber \\
&& \;\;\;\;\;\;\;\;\;\;\;\;\;\;\;\;\;\;\;\;\;
         \left .
              + t_j \left \{
                f^{\dagger}_{(j+1) n \sigma} f_{j n \sigma}
              + {\rm H.c.} \right \}
         \right ]
\nonumber \\
&+&      \frac{J_{\perp}}{2} \sum_{n = 1}^{k}
         \left [
              f^{\dagger}_{0n\downarrow}f_{0n\uparrow}S^+
              + {\rm H.c.}
         \right ]
\nonumber \\
&+&      \frac{J_z}{2} \sum_{n = 1}^{k}
         S_z\! \left [
              f^{\dagger}_{0 n \uparrow} f_{0 n \uparrow}
              - f^{\dagger}_{0 n \downarrow} f_{0 n \downarrow}
         \right ]
         - g_i h S_z
\nonumber \\
&-& \frac{g_e h}{2}
         \sum_{j = 0}^{\infty}
         \sum_{n = 1}^{k}
         \left [
              f^{\dagger}_{j n \uparrow} f_{j n \uparrow}
              - f^{\dagger}_{jn\downarrow} f_{jn\downarrow}
         \right ] .
\label{H-lattice}
\end{eqnarray}
Here $\epsilon_j$ and $t_j$, respectively, are the on-site
energies and hopping matrix elements along the chain. Any
lattice model with identical noninteracting bands can be
cast in the form of Eq.~(\ref{H-lattice}) using a Wilson-type
construction.~\cite{Wilson75} Different lattice models are
distinguished by the tight-binding parameters $\epsilon_j$
and $t_j$, the largest of which determines the bandwidth
$D$. For example, particle-hole symmetry demands that
$\epsilon_j = 0$ for all sites along the chain. In the
following we assume a large longitudinal coupling,
$J_z \gg D, |J_{\perp}|$, and expand in powers of $1/J_z$
to derive an effective low-energy Hamiltonian at energies
far below $J_z$.

For $k_B T \ll J_z$, the fermionic degrees of freedom at
site zero bind tightly to the impurity so as to minimize
the $J_z$ interaction term. There are two degenerate
ground states of this interaction term:
\begin{equation}
| + \rangle =
       \prod_{n = 1}^{k} f^{\dagger}_{0 n \uparrow}
                       | S_z = \downarrow \rangle \; ,
\;\;\;\;\;
| - \rangle =
       \prod_{n = 1}^{k} f^{\dagger}_{0 n \downarrow}
                       | S_z = \uparrow \rangle \; ,
\end{equation}
corresponding to the total spin projections
$S_{\rm total}^z = \pm (k-1)/2$. All excited eigenstates
of the $J_z$ interaction term are thermally inaccessible,
being removed in energy by integer multiples of $J_z/4$.
Defining a new isospin operator $\vec{\tau}$ according to
\begin{equation}
\tau_z = \frac{1}{2}
         \sum_{p = \pm} p |p\rangle \langle p| \; ,
\;\;\;\;\;
\tau^{\pm} = |\pm\rangle \langle\mp|
\label{tau-def}
\end{equation}
[compare with Eqs.~(\ref{tau-z-def})--(\ref{tau-perp-def})
and Eqs.~(\ref{tau-z-def2})--(\ref{tau-perp-def2})],
the effective low-energy Hamiltonian takes the form
${\cal H}_{\rm chain} + {\cal H}_{\rm mag} +
{\cal H}_{\rm int}$, where ${\cal H}_{\rm chain}$ is
the tight-binding Hamiltonian of the truncated chain
with site $j = 0$ removed, ${\cal H}_{\rm mag}$ is the
magnetic-field term
\begin{equation}
{\cal H}_{\rm mag} =
         -g_{\tau} h \tau_z - \frac{g_e h}{2}
         \sum_{j = 1}^{\infty}
         \sum_{n = 1}^{k}
         \left [
              f^{\dagger}_{j n \uparrow} f_{j n \uparrow}
              - f^{\dagger}_{j n\downarrow} f_{j n\downarrow}
         \right ]
\label{H-mag}
\end{equation}
with $g_{\tau} = k g_e - g_i$, and ${\cal H}_{\rm int}$
contains all finite-order corrections in either $1/J_z$
or $J_{\perp}$ (or both).

The explicit form of the Hamiltonian term ${\cal H}_{\rm int}$
depends on the number of conduction-electron channels $k$.
For $k = 1$ and up to linear order in $1/J_z$, it takes
the form
\begin{equation}
{\cal H}_{\rm int} = J_{\perp} \tau_x
        + \frac{\lambda_z}{2} \tau_z
        \left [
             f^{\dagger}_{1 \uparrow} f_{1 \uparrow}
           - f^{\dagger}_{1 \downarrow} f_{1 \downarrow}
         \right ] ,
\end{equation}
where $\lambda_z$ equals
\begin{equation}
\lambda_z = \frac{4 t_0^2}{J_z} \ll D .
\label{lambda_z}
\end{equation}
Here we have omitted the redundant channel index $n$
from within $f^{\dagger}_{1 n \sigma}$, i.e., we have set
$f^{\dagger}_{1 n \sigma} \to f^{\dagger}_{1 \sigma}$.
Since the $\lambda_z$ term is exactly marginal, the
low-temperature physics is governed for $h = 0$ by
the spin-flip term $J_{\perp}$, which favors the
singlet state $\frac{1}{\sqrt{2}} \left[ |+\rangle
- |-\rangle \right]$ for $J_{\perp} > 0$. In this
manner one recovers the characteristic spin singlet
of the single-channel Kondo effect, restoring thereby
SU(2) spin symmetry as $T \to 0$. A local magnetic field
breaks the emerging SU(2) spin symmetry as it physically
should by introducing weak spin-dependent scattering
at the open end of the truncated chain.

The situation is somewhat different for $k = 2$. In this
case ${\cal H}_{\rm int}$ acquires the form
\begin{eqnarray}
{\cal H}_{\rm int} &=&
        \frac{\lambda_z}{2} \tau_z \sum_{n = 1}^{2}
        \left [
             f^{\dagger}_{1 n\uparrow} f_{1 n\uparrow}
           - f^{\dagger}_{1 n\downarrow} f_{1 n\downarrow}
         \right ]
\nonumber \\
&+& \lambda_{\perp} \sum_{n = 1}^{2}
        \left [
             f^{\dagger}_{1 n\uparrow} f_{1 n\downarrow} \tau^-
             + {\rm H.c.}
         \right ] ,
\label{H_int-2CK}
\end{eqnarray}
where $\lambda_z$ is still given to order $1/J_z$ by
Eq.~(\ref{lambda_z}), and $\lambda_{\perp}$ scales as
\begin{equation}
\lambda_{\perp} \sim \frac{ J_{\perp} t_0^2}{J_z^2} .
\end{equation}
Here we have omitted higher order interaction terms within
${\cal H}_{\rm int}$, and restricted attention to
particle-hole symmetry. Away from particle-hole symmetry,
an additional potential-scattering term of the form
$\Delta \epsilon_1 \sum_n f^{\dagger}_{1 n \sigma}
f_{1 n \sigma}$ is generated at second order in $1/J_z$.

Equation~(\ref{H_int-2CK}) has the same exact form as the
original spin-exchange interaction in Eq.~(\ref{H-lattice}),
but with two modifications: (i) The coupling constants
$J_z$ and $J_{\perp}$ have been pushed to weak coupling;
(ii) The site $j = 0$ has been replaced with $j = 1$. The
physical content of the local moment $\vec{\tau}$ has also
changed. Importantly, since
$D \gg \lambda_z \gg |\lambda_{\perp}|$ one lies within
the confines of the antiferromagnetic domain, rendering
${\cal H}_{\rm int}$ a relevant perturbation. Hence,
in accordance with the bosonization treatment of
Sec.~\ref{sec:mapping}, the strong-coupling limit $J_z \gg D$
maps onto weak coupling, extending the duality of Nozi\'eres
and Blandin~\cite{NB80} to large spin-exchange anisotropy.
Moreover, since $t_0 \sim 1/\rho$ for conventional lattice
models, then $\lambda_z$ of Eq.~(\ref{lambda_z}) is
comparable to $J'_z$ of Eq.~(\ref{2chK}). The main
difference as compared to bosonization pertains to the
transverse coupling $J_{\perp}$, which renormalizes to
$\lambda_{\perp} \propto 1/J_z^2$ in the strong-coupling
expansion, but is left unchanged within bosonization.
Excluding this rather minor discrepancy, the two approaches
are in close agreement with one another.

The crucial difference for $k > 2$ has to do with the
dynamic components of ${\cal H}_{\rm int}$, responsible
for flipping the isospin $\vec{\tau}$. To see this we
note that the total spin projections $S_{\rm total}^z$
of the states $|+\rangle$ and $|-\rangle$ differ by
$k - 1$. Since the Hamiltonian of Eq.~(\ref{H-lattice})
preserves the overall spin projection of the entire system,
then the flipping of $\tau_z$ from plus to minus or vice
versa must be accompanied by an opposite spin flip of
$k - 1$ electrons along the truncated chain. Hence the
leading dynamic term in ${\cal H}_{\rm int}$ only shows
up at order $(1/J_z)^{2(k-1)}$, taking the form
\begin{equation}
\lambda_{\perp} \left [
                \tau^{-} \sum_{m = 1}^{k}
                \prod_{n \neq m}
                f^{\dagger}_{1 n\uparrow} f_{1 n\downarrow}
                + {\rm H.c.} \right ]
\label{lambda_p-k}
\end{equation}
with
\begin{equation}
\lambda_{\perp} \sim J_{\perp} \left (
                \frac{t_0}{J_z} \right )^{2(k-1)} .
\end{equation}
In contrast to the dynamic part, the leading static
component of ${\cal H}_{\rm int}$ remains given by the
$\lambda_z$ term of Eq.~(\ref{H_int-2CK}), except for the
summation over $n$ which now runs over all $k$ channels.
To linear order in $1/J_z$ the coupling $\lambda_z$ is
independent of $k$, being given by Eq.~(\ref{lambda_z}).
Once again, an additional potential-scattering term is
generated at order $(1/J_z)^2$ away from particle-hole
symmetry.

Since the $\lambda_{\perp}$ term of Eq.~(\ref{lambda_p-k})
involves the creation and annihilation of $k-1$ electrons
at the open end of the truncated chain, it has the scaling
dimension $\Delta_{\perp} = k-1$ with respect to the
$J_z \to \infty$ ``free'' Hamiltonian. Hence this term is
irrelevant for $k > 2$. The same holds true of all higher
order dynamical terms generated, as these likewise contain
at least $k-1$ creation and $k-1$ annihilation operators
of electrons localized along the truncated chain. Similar
to the case $s > 1/2$ and $|J_z| > |J_{\perp}|$ in
Sec.~\ref{sec:Perturbative-RG}, the resulting fixed-point
Hamiltonian for $h = 0$ corresponds then to a finite
longitudinal coupling $\lambda_z$ but zero
$\lambda_{\perp}$, in perfect agreement with the results
of bosonization. In fact, an identical scaling dimension
$\Delta_{\perp} \to k -1$ is obtained in the Anderson-Yuval
approach for $\rho J_z \gg 1$,~\cite{comment_on_AY}
reinforcing the qualitative agreement between the two
approaches. Most importantly, a line of stable
ferromagnetic-like fixed points is seen to exist for
any lattice model with sufficiently large $J_z > 0$,
just as predicted by bosonization.

Evidently, the strong-coupling expansion confirms the
results of bosonization for all three cases: $k = 1$,
$k = 2$, and $k > 2$. Moreover, it provides a transparent
physical picture for the source of distinction between
the three cases. It all boils down to the nature of the
local spin configurations selected by a sufficiently large
$J_z > 0$. We therefore conclude that Fig.~\ref{fig:fig1}(a)
correctly describes the phase diagram of the
spin-$\frac{1}{2}$ multichannel Kondo model with $k > 2$
conduction-electron channels.

\section{Strong-coupling expansion for arbitrary spin $s$}
\label{sec:SC-general-s}

Our discussion in the previous two sections was confined
to a spin-$\frac{1}{2}$ impurity. An appealing feature of
the strong-coupling expansion in $1/J_z$ is that it can
easily be generalized to arbitrary spin $s$. This is the
goal of the present section.

The basic considerations for arbitrary $s > 0$ are quite
similar to those for $s = 1/2$. As before, the $J_z$
interaction term possesses two degenerate ground states.
We label these states according to
\begin{equation}
| + \rangle = \prod_{n = 1}^{k} f^{\dagger}_{0 n \uparrow}
              | S_z = -s \rangle \; ,
\;\;\;\;\;
| - \rangle = \prod_{n = 1}^{k} f^{\dagger}_{0 n \downarrow}
              | S_z = s \rangle
\label{|pm>}
\end{equation}
for $k \geq 2s$, and
\begin{equation}
| + \rangle = \prod_{n = 1}^{k} f^{\dagger}_{0 n \downarrow}
              | S_z = s \rangle \; ,
\;\;\;\;\;
| - \rangle = \prod_{n = 1}^{k} f^{\dagger}_{0 n \uparrow}
              | S_z = -s \rangle
\end{equation}
for $k < 2s$. With this convention, $|\pm\rangle$ has the
total spin projection $S_{\rm total}^z = \pm |k/2 - s|$.
Defining the isospin $\vec{\tau}$ according to
Eq.~(\ref{tau-def}), the effective low-energy Hamiltonian
is written as ${\cal H}_{\rm chain} + {\cal H}_{\rm mag} +
{\cal H}_{\rm int}$, where ${\cal H}_{\rm chain}$ is
the tight-binding Hamiltonian of the truncated chain
with site $j = 0$ removed, ${\cal H}_{\rm mag}$ is
the magnetic-field term of Eq.~(\ref{H-mag}), and
${\cal H}_{\rm int}$ contains all finite-order corrections
in either $1/J_z$ or $J_{\perp}$ (or both). The sole
modification to ${\cal H}_{\rm mag}$ of Eq.~(\ref{H-mag})
is in the effective $g$-factor $g_{\tau}$, which changes
from $g_e k - g_i$ for $s = 1/2$ to $\pm(g_e k - 2s g_i)$
for arbitrary $s$. Here the plus (minus) sign corresponds
to $k \ge 2s$ ($k < 2s$).

Similar to the case $s = 1/2$, the delicate interplay
between $k$ and $s$ enters through the Hamiltonian term
${\cal H}_{\rm int}$. Let us separate the discussion of
the dynamic and static components of ${\cal H}_{\rm int}$,
as these depend differently on $s$ and $k$. The leading
static component of ${\cal H}_{\rm int}$ remains given
by the $\lambda_z$ term of Eq.~(\ref{H_int-2CK}), except
for the summation over $n$ which now runs over all $k$
conduction-electron channels. The coupling $\lambda_z$
does depend on $k - 2s$, however only through its
sign. For $k \geq 2s$ it is given to order $1/J_z$ by
Eq.~(\ref{lambda_z}), corresponding to an antiferromagnetic
interaction. For $k < 2s$ it acquires an additional minus
sign, corresponding to a ferromagnetic interaction. Apart
from its overall sign, the leading static component of
${\cal H}_{\rm int}$ is independent of $s$.

Moving on to the dynamic part of ${\cal H}_{\rm int}$,
its leading-order component displays a more elaborate
dependence on $k$ and $s$. As noted above, the total spin
projections of the states $|+\rangle$
and $|-\rangle$ differ by $|k - 2s|$. Consequently, the
flipping of $\tau_z$ from up to down or vice versa must
be accompanied by an opposite spin flip of $|k - 2s|$
electrons along the truncated chain, else the total
spin projection of the system is not conserved. This
consideration dictates the following form for the
leading dynamical term:
\begin{equation}
\lambda_{\perp} \left [
        \tau^- \hat{O}^{+}_{k,s} + \tau^+ \hat{O}^{-}_{k,s}
        \right ] ,
\end{equation}
where $\hat{O}^{+}_{k,s} = ( \hat{O}^{-}_{k,s} )^{\dagger}$
is a channel-symmetric operator that creates $|k - 2s|$
spin-up electrons and annihilates $|k - 2s|$ spin-down
electrons at, or close as possible to, the open end of
the truncated chain.

The operator $\hat{O}^{+}_{k,s}$ is generally too
complicated to write down. It greatly simplifies in two
cases. For $k = 2s$, $\hat{O}^{+}_{k,s}$ reduces to
the unity operator; For $s = 1/2$ and $k > 1$, it is
given by Eq.~(\ref{lambda_p-k}). As for the coupling
$\lambda_{\perp}$, it scales differently for
$k \ge s$ and $k < s$. For $k \ge s$, $\lambda_{\perp}$
behaves as
\begin{equation}
\lambda_{\perp} \sim J_{\perp}
        \left ( \frac{ J_{\perp} }{ J_z } \right )^{2s - 1}
        \left ( \frac{t_0}{J_z} \right )^{2|k - 2s|} .
\end{equation}
For $k < s$, it involves a higher power of $1/J_z$, as
electrons farther into the chain must participate in
the flipping of $\tau_z$. With the possible exception
of $k = 2s$, the transverse coupling $|\lambda_{\perp}|$
is parametrically smaller than $|\lambda_z|$, a fact
that will have important implications later on.

As a function of $s$ and $k$, $\lambda_{\perp}$ has the
scaling dimension $\Delta_{\perp} = |k - 2s|$ with respect
to the $J_z \to \infty$ ``free'' Hamiltonian. For $h = 0$,
this yields the following classification of the low-energy
physics.

(i) $s = k/2$ --- In this exactly screened case,
$\lambda_{\perp}$ acts as a local transverse magnetic
field, lifting the two-fold degeneracy of $|\pm\rangle$.
Although the local state favored by $\lambda_{\perp}$ is
generally not an SU(2) spin singlet, the low-energy physics
is identical to that for isotropic antiferromagnetic
exchange, as can be seen from the boundary
condition imposed on the truncated chain. Specifically,
a local Fermi liquid progressively forms below the Kondo
temperature $T_K \sim \lambda_{\perp}$.

(ii) $s = k/2 - 1/2$ --- In this overscreened case, the
Hamiltonian term ${\cal H}_{\rm int}$ has the same exact form
as Eq.~(\ref{H_int-2CK}), except for the summation over $n$
which now runs over all $k$ conduction-electron channels.
Hence the system is described by a weakly coupled $k$-channel
Kondo Hamiltonian with $\lambda_z \gg |\lambda_{\perp}| > 0$.
The effect of the large spin-exchange anisotropy in the
original Hamiltonian is to reduce the effective impurity
moment from $s$ to $1/2$. Based on the perturbative RG
analysis of Sec.~\ref{sec:Perturbative-RG}, the resulting
Hamiltonian flows to the overscreened fixed point of
the $k$-channel, spin-$\frac{1}{2}$ Kondo Hamiltonian,
which is equivalent in turn to that of
the $k$-channel Kondo model with $s = k/2 - 1/2$.
As in the exactly screened case discussed above, a
large spin-exchange anisotropy $J_z \gg D, |J_{\perp}|$
is seen to be irrelevant for $s = k/2 - 1/2$.

(iii) $s = k/2 + 1/2$ --- Similar to the previous
case, the system is described by a weakly coupled
$k$-channel Kondo Hamiltonian with $s \to 1/2$.
However, the effective coupling is now ferromagnetic:
$-\lambda_z \gg |\lambda_{\perp}| > 0$. Consequently,
the flow is to a line of stable ferromagnetic-like fixed
points with finite $\lambda_z$ but zero $\lambda_{\perp}$.
The resulting low-energy physics is that of a singular
Fermi liquid~\cite{Mehta-etal05} plus a residual isospin
$\vec{\tau}$. It differs from that of an isotropic
spin-exchange interaction only in the residual
$\lambda_z$ interaction. Since the latter term is
marginal, so is the large spin-exchange anisotropy
for this underscreened case.

\begin{figure}
\centerline{
\includegraphics[width=80mm]{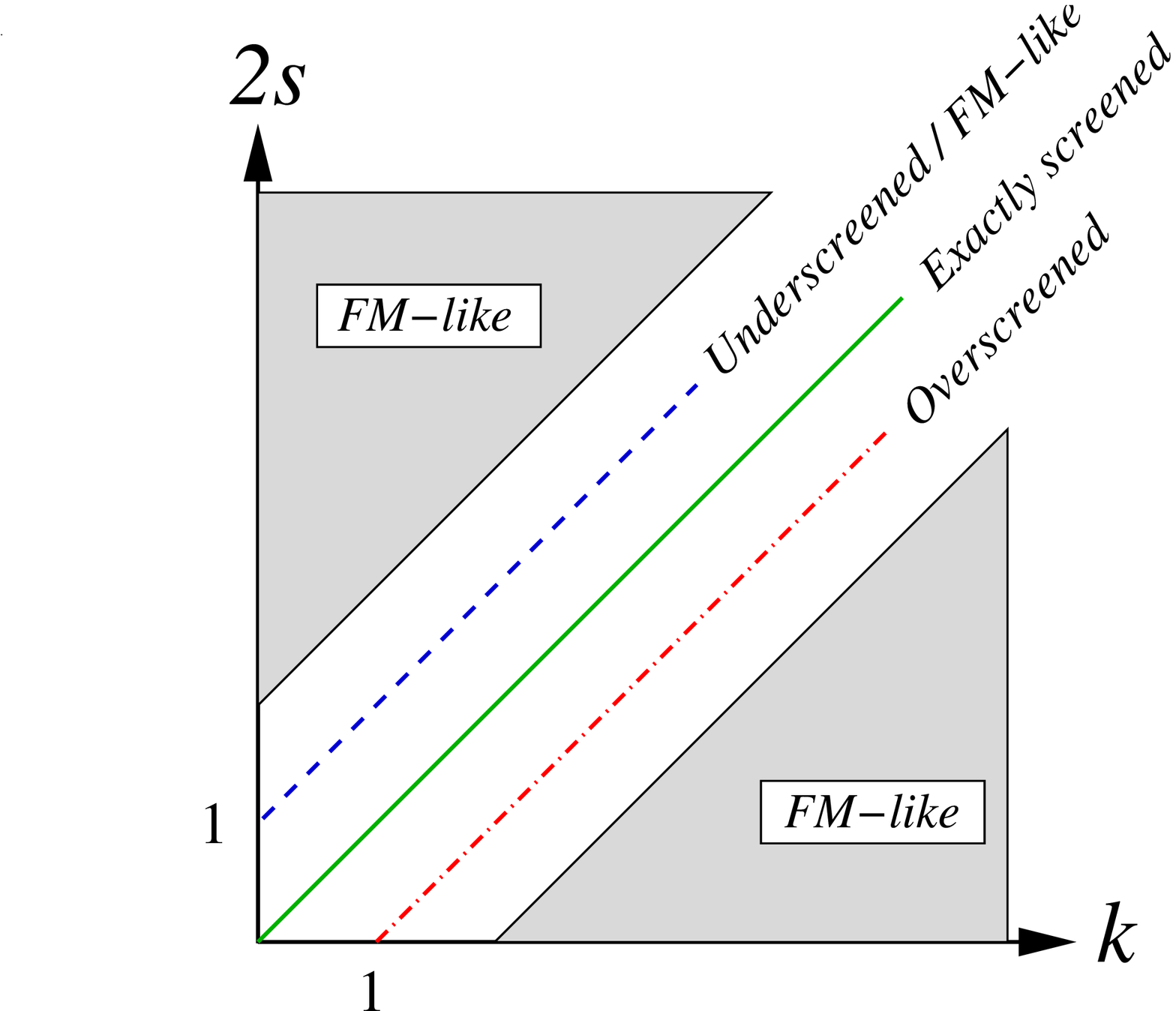}
}\vspace{-5pt}
\caption{(Color online)
         Phase diagram of the multichannel Kondo model with
         $J_z \gg |J_{\perp}|, D$, as a function of $k$ and
         $s$. For $2s = k$ and $2s = k - 1$, the low-energy
         physics is stable against such a large spin-exchange
         anisotropy. The fixed point remains that of an
         exactly screened ($2s = k$) or an overscreened
         ($2s = k - 1$) Kondo effect. For $|k - 2s| > 1$, the
         low-energy physics is unstable against such a large
         spin-exchange anisotropy. The system flows to a
         line of stable ferromagnetic-like fixed points with
         a residual isospin-$\frac{1}{2}$ local moment.
         For $2s = k + 1$, a large spin-exchange anisotropy
         is marginal. The system still flows to a line of
         stable ferromagnetic-like fixed points, however
         these differ from the isotropic underscreened fixed
         point only by the marginal operator $\lambda_z$
         (see text).}
\label{fig:fig2}
\end{figure}

(iv) $|k - 2s| > 1$ --- In this effectively underscreened
case, the scaling dimension $\Delta_{\perp}$ exceeds one.
Hence $\lambda_{\perp}$ is irrelevant, as are all higher
order dynamical terms generated within ${\cal H}_{\rm int}$.
The system thus flows to a line of stable ferromagnetic-like
fixed points, characterized by a finite $\lambda_z$ but zero
$\lambda_{\perp}$. The low-energy physics is again that
of a (potentially singular) Fermi liquid plus a residual
isospin $\vec{\tau}$. For $s < k/2 - 1/2$, this differs
markedly from the overscreened fixed point of the
isotropic spin-exchange model. Also for $s > k/2 + 1/2$
this differs from the underscreened fixed point of
the isotropic model, as the residual local-moment
degeneracy is two instead of $2s - k + 1 > 2$. Importantly,
in all cases the resulting low-energy physics is insensitive
to the sign of $k - 2s$, in stark contrast to the isotropic
case. A large spin-exchange anisotropy of the form
$J_z \gg D, |J_{\perp}|$ is therefore relevant with
respect to the isotropic fixed point for all
$|k - 2s| > 1$.

The phase diagram of the multichannel Kondo Hamiltonian
with $J_z \gg |J_{\perp}|, D$ is summarized in
Fig.~\ref{fig:fig2} as a function of $k$ and $s$.
Excluding the exactly screened line $2s = k$ and
the overscreened line $2s = k - 1$, the model
flows to a line of stable ferromagnetic-like fixed
points with a residual isospin-$\frac{1}{2}$ local
moment for all $k$ and $s$. For $2s = k + 1$, the
ferromagnetic-like fixed points and the isotropic
underscreened fixed point are equivalent. They only
differ by the marginal operator $\lambda_z$. For
$2s > k + 1$, the ferromagnetic-like fixed points and
the isotropic underscreened fixed point are distinctly
different, possessing a different residual
degeneracy.

\section{NRG study of $s = 1/2$, $k = 3$}
\label{sec:NRG}

Although the strong-coupling expansion unequivocally
confirms the existence of a line of stable
ferromagnetic-like fixed points for large $J_z$ and
$|k - 2s| > 1$, it cannot access the entire phase
diagram of the anisotropic multichannel Kondo Hamiltonian.
In particular, the second Kosterliz-Thouless line
predicted by bosonization for $s = 1/2$ and $k > 2$
[see Fig.~\ref{fig:fig1}(a)] lies beyond the scope of
this approach. In this section, we use Wilson's
numerical renormalization-group (NRG)
method~\cite{Wilson75} to conduct a systematic study
the phase diagram of the anisotropic multichannel
Kondo Hamiltonian with $s = 1/2$ and $k = 3$.

In conventional formulations of the NRG,~\cite{Wilson75}
one considers a particular choice for the tight-binding
parameters in Eq.~(\ref{H-lattice}), given by
$\epsilon_j = 0$ and
$t_j = D_{\Lambda} \Lambda^{-j/2} \xi_j$ with
\begin{equation}
D_{\Lambda} = \frac{D}{2}(1 + \Lambda^{-1}) ,
\end{equation}
\begin{equation}
\xi_{j} = \frac{1-\Lambda^{-(j+1)}}
    {\sqrt{(1-\Lambda^{-(2j+1)})(1-\Lambda^{-(2j+3)})}} .
\end{equation}
Here $\Lambda > 1$ is a discretization parameter. For
$\Lambda \to 1^+$, the resulting model describes an
impurity spin locally coupled to $k$ identical conduction
bands with a symmetric box density of states
$\rho(E) = (1/2D) \theta (D - |E|)$. For $\Lambda > 1$,
the model represents an impurity spin
coupled to a logarithmically discretized version
of the same bands.~\cite{Wilson75}

The first step in the NRG approach is to recast the 
Hamiltonian ${\cal H}$ as the limit of a sequence of
dimensionless Hamiltonians ${\cal H}_N$:
\begin{equation}
{\cal H} = \lim_{N \rightarrow \infty}
           \left\{
                 D_{\Lambda} \Lambda^{-(N-1)/2} {\cal H}_{N}
           \right\}
\end{equation}
with
\begin{eqnarray}
{\cal H}_{N} &=& \Lambda^{(N-1)/2}
         \left [
         \frac{J_{\perp}}{2 D_{\Lambda}} \sum_{n = 1}^{k}
         \left (
              f^{\dagger}_{0n\downarrow}f_{0n\uparrow}S^+
              + {\rm H.c.}
         \right )
         \right .
\label{H-NRG} \\
&+&
         \frac{J_z}{2 D_{\Lambda}} \sum_{n = 1}^{k}
         S_z \left (
              f^{\dagger}_{0 n \uparrow} f_{0 n \uparrow}
              - f^{\dagger}_{0 n \downarrow} f_{0 n \downarrow}
         \right )
\nonumber \\
&+& \left .
         \sum_{j = 0}^{N - 1}
         \sum_{n = 1}^{k}
         \sum_{\sigma = \uparrow, \downarrow}
              \Lambda^{-j/2} \xi_j
              \left \{
              f^{\dagger}_{(j+1) n \sigma} f_{j n \sigma}
              + {\rm H.c.} \right \}
         \right ] .
\nonumber
\end{eqnarray}
The finite-size Hamiltonians ${\cal H}_N$ are then
diagonalized iteratively using the NRG transformation
\begin{equation}
{\cal H}_{N+1} = \sqrt{\Lambda}{\cal H}_N +
               \sum_{n = 1}^{k}
               \sum_{\sigma = \uparrow, \downarrow}
               \xi_{N}
               \left \{
                       f^{\dagger}_{(N+1) n \sigma}
                       f_{N n \sigma} + {\rm H.c.}
               \right \} .
\label{NRG_iteration}
\end{equation}
Here the prefactor $\Lambda^{(N-1)/2}$ in Eq.~(\ref{H-NRG})
guarantees that the low-energy excitations of ${\cal H}_N$
are of order one for all $N$. The approach to a fixed-point
Hamiltonian is signaled by a limit cycle of the NRG
transformation, with ${\cal H}_{N + 2}$ and ${\cal H}_N$
sharing the same low-energy spectrum.

It practice, it is impossible to keep track of the
exponential increase in the size of the Hilbert space as
a function of the chain length $N$. Hence only the lowest
$N_s$ eigenstates of ${\cal H}_N$ are retained at the
conclusion of each iteration.~\cite{comment_on_trunctaion}
The retained states are used
in turn to construct the eigenstates of ${\cal H}_{N+1}$.
The truncation error involved can be systematically
controlled by varying $N_s$. However, since the size of
the Hilbert space increases by a factor of $2^{2k}$ with
each additional iteration, only a moderately small number
of conduction-electron channels can be treated reliably
using present-day computers, typically no more than $3$
or $4$ channels. In the following we focus on $k = 3$
and $s = 1/2$, which is the simplest version of the
multichannel Kondo Hamiltonian that is both amendable
to the NRG and predicted to display the phase diagram
of Fig.~\ref{fig:fig1}(a).

To cope with the large computational effort involved in
exploring the phase diagram of the three-channel Kondo
model with spin-exchange anisotropy, we used rather large
values of $\Lambda$ (either $\Lambda = 3$ or $\Lambda = 5$),
and applied an alternative truncation scheme to the one
customarily used. Since each of the NRG Hamiltonians
${\cal H}_N$ is block diagonal in the conserved quantum
numbers,~\cite{comment_on_conserved_numbers} the
computational time is mostly governed by the largest
block to be diagonalized. Thus, instead of retaining
a fixed number of eigenstates of ${\cal H}_N$, at each
iteration we adjusted the threshold energy for truncation
so that the largest block did not exceed $N_{\rm block}$
states.~\cite{comment_on_trunctaion} For $\Lambda = 3$
we used $N_{\rm block} = 100$, while
for $\Lambda = 5$ we set $N_{\rm block} = 80$. This
approach gave rise to variations in the number of
states retained at the conclusion of each iteration.
Typically, 3000 to 4000 states were retained.
In some extreme cases down to 2500 states and up to
5500 states were kept. We have verified that the
threshold energies selected in this way were sufficiently
high so as not to spoil the accuracy of the calculations.

\begin{widetext}

\begin{figure}[h]
\centerline{
\includegraphics[width=165mm]{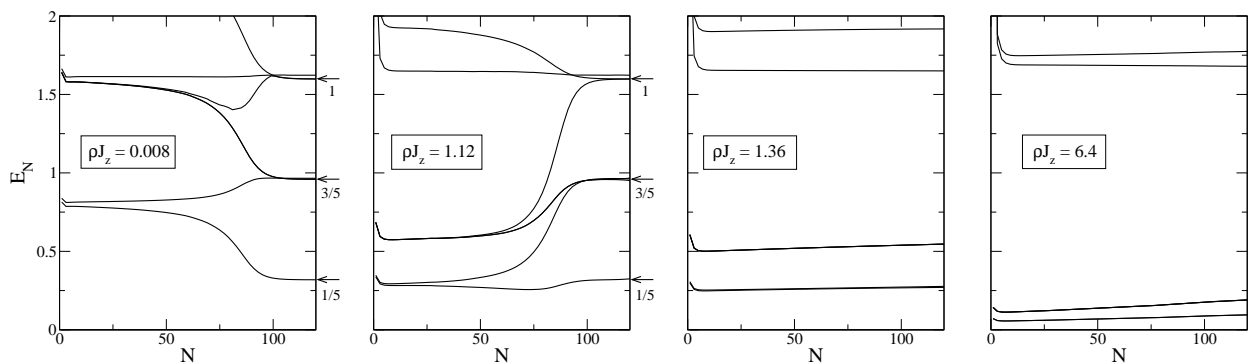}
}\vspace{-5pt}
\caption{NRG level flow (odd iterations) for $s = 1/2$,
         $k = 3$, $\rho J_{\perp} = 0.008$, and increasing
         values of $\rho J_z$. Here $\Lambda = 3$ was used.
         For $\rho J_z < \rho J_z^c \approx 1.26$ (left two
         panels), the system flows
         to the non-Fermi-liquid fixed point of the
         three-channel Kondo effect, irrespective of $J_z$.
         There is excellent agreement with the finite-size
         spectrum of three-channel Kondo effect obtained
         from conformal field theory, whose energy levels
         (in units of the fundamental level spacing)
         are indicated by arrows. For $J_z > J_z^c$ (right
         two panels), the finite-size spectrum remains
         Fermi-liquid-like down to minuscule energies,
         without crossing over to the non-Fermi-liquid
         fixed point of the three-channel Kondo effect. The
         persisting drift of NRG levels for $J_z > J_z^c$
         appears to be due to some weak numerical
         instability. A similar drift of levels occurs for
         $-J_z \gg J_{\perp} > 0$ (ferromagnetic coupling).}
\label{fig:fig3}
\end{figure}

\end{widetext}

Figure~\ref{fig:fig3} depicts the finite-size spectra
obtained for $\rho J_{\perp} = 0.008$ and different
values of $\rho J_z > 0$. Here $\Lambda = 3$ was used.
Up to a critical coupling $\rho J_z^{c} \approx 1.26$,
the system always flows to the non-Fermi-liquid fixed
point of the three-channel Kondo effect. Indeed, the
fixed-point spectrum is independent of $J_z < J_z^c$,
and is identical to that for isotropic spin-exchange
interaction (left-most panel). There is excellent agreement
with the finite-size spectrum of the three-channel Kondo
effect obtained from conformal field theory,~\cite{AL91}
whose energy levels are marked by arrows in
Fig.~\ref{fig:fig3}. Here the slight splitting of NRG
levels near the dimensionless energy $1.6$ is a
$\Lambda$-dependent feature. This splitting is reduced
upon decreasing $\Lambda$, and should disappear for
$\Lambda \to 1^+$.

A different picture is recovered for $J_z > J_z^{c}$.
(i) As demonstrated in Fig.~\ref{fig:fig3} for
    $\rho J_z = 1.36$ and $\rho J_z = 6.4$, the
    finite-size spectrum remains Fermi-liquid-like down
    to the lowest energies reached ($E \sim 10^{-120}D$
    in some extreme runs), without crossing over to the
    overscreened fixed point of the three-channel Kondo
    effect.
(ii) The finite-size spectrum varies continuously with
     $J_z > J_z^c$, suggesting the flow to a line of
     fixed points connected by a marginal operator.
(iii) For $J_z \gg J_z^c$, the quantum numbers and
      degeneracies of the NRG levels are in excellent
      agreement with those anticipated based on the
      strong-coupling analysis of Sec.~\ref{sec:SC-s=1/2}.
(iv) The NRG level flow confirms that channel asymmetry is
     a marginal perturbation for $J_z > J_z^c$ (not shown),
     in stark contrast to the regime $J_z < J_z^c$.
These features are all consistent with the flow to
a line of stable ferromagnetic-like fixed points as
predicted by bosonization.

We do note, however, a persisting drift of the NRG levels
for $J_z > J_z^c$. A similar drift of levels occurred for
$-J_z \gg |J_{\perp}| > 0$ (ferromagnetic coupling), and
appears to be driven by some weak numerical instability.
While we cannot entirely rule out an eventual crossover
to the non-Fermi-liquid fixed point of the three-channel
Kondo effect at some lower temperature, we find this
scenario highly unlikely given the extremely low energy
scales reached and the rapid change of behavior as $J_z$
is swept through $J_z^c$. We therefore identify $J_z^{c}$
with the phase boundary between the antiferromagnetic
and the new ferromagnetic-like domain predicted by
bosonization.

\begin{figure}[tb]
\centerline{
\includegraphics[width=75mm]{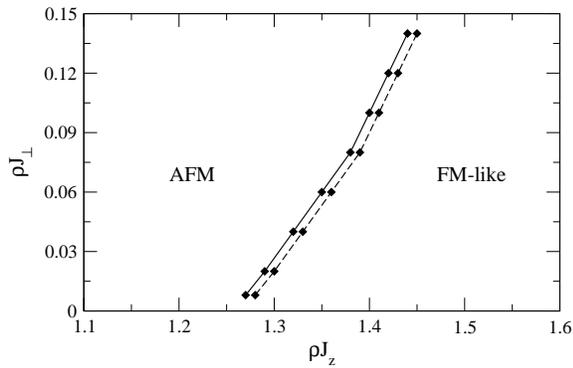}
}\vspace{-5pt}
\caption{NRG phase boundary between the antiferromagnetic
         and the new ferromagnetic-like domain, for
         $s = 1/2$, $k = 3$ and $\Lambda = 5$. Points on
         the solid line all converged to the non-Fermi-liquid
         fixed point of the three-channel Kondo effect. Points
         on the dashed line showed no such tendency up
         to $N = 120$ NRG iterations, corresponding to
         $E \sim 10^{-42} D$. We estimate the phase boundary
         between the two domains to lie in between the solid
         and dashed lines.}
\label{fig:fig4}
\end{figure}

Guided by this interpretation, we turned to explore the
dependence of $J_z^c$ on $J_{\perp}$. The resulting
phase diagram is plotted in Fig.~\ref{fig:fig4}, for
$\Lambda = 5$. While points on the solid line all
converged to the non-Fermi-liquid fixed point of the
three-channel Kondo effect, points on the dashed line
showed no such tendency up to $N = 120$ NRG iterations
(corresponding to $E \sim 10^{-42} D$). We therefore
estimate the phase boundary between the antiferromagnetic
and the ferromagnetic-like domains to lie in between the
solid and dashed lines. In contrast to the solid line,
which definitely lies on the antiferromagnetic side
of the transition, one
cannot guarantee that all points on the dashed line
lie on the ferromagnetic-like side. We have confirmed,
however, that the phase diagram of Fig.~\ref{fig:fig4} is
practically unchanged upon going to $N = 180$ iterations.
We further stress that the exact position of the phase
boundary is in general $\Lambda$ dependent, although
its dependence on $\Lambda$ appears to be weak. For
$\Lambda = 3$, for example, the position of the phase
boundary varied by no more than just a few percent. We
expect a similar proximity of the $\Lambda \to 1^+$
phase boundary.

The above results are clearly in good qualitative agreement
with the bosonization treatment of Sec.~\ref{sec:mapping}.
We now turn to a more quantitative comparison. Setting
$k = 3$ in Eq.~(\ref{J-ast}), the critical coupling
$J_z^{\ast}$ predicted by bosonization is equal to
$\rho J_z^{\ast} = 4\sqrt{3}/\pi \approx 2.2$. Based on the
NRG results for $\Lambda = 3$ and $\Lambda = 5$, we estimate
the location of the critical coupling for a symmetric box
density of states (i.e., for $\Lambda \to 1^+$) to be around
$\rho J_z^{\ast} \sim 1.2 - 1.3$. Thus, the bosonization and
NRG results are within a factor of two from one another.
Considering that $J_z^{\ast}$ lies well beyond the strict
range of validity of bosonization, we find this degree of
agreement to be quite remarkable.
As for the shape of the phase boundary, it was predicted
in Sec.~\ref{sec:mapping} to have the linear form $J_z =
J_z^{\ast} + C_k |J_{\perp}|$, where $0 < C_k < 1$ is given
by Eq.~(\ref{C_k}). As seen in Fig.~\ref{fig:fig4}, the NRG
phase boundary for $\Lambda = 5$ is well described by a
linear curve, at least up to $\rho J_{\perp} \approx 0.08$.
The corresponding slope $C_{\rm NRG} \approx 0.67$ is
indeed less than one, but is nearly three-fold larger
than the bosonization result, $C_{k=3} = 0.25$. This
discrepancy can be largely accounted for by plugging
the NRG value for $\rho J_z^{\ast}$ into the right-hand
side of Eq.~(\ref{C_k}), which yields $C \approx 0.5$.

\section{Discussion and summary}
\label{sec:discussion}

The multichannel Kondo Hamiltonian is an important
paradigm in correlated electron systems, with
possible applications to varied systems. Depending on
the size of the impurity spin, $s$, and the number
of independent conduction-electron channels, $k$,
it can display either local Fermi-liquid, singular
Fermi-liquid, or non-Fermi-liquid behavior. Although
the isotropic model is by now well understood, we
have shown in this paper that an XXZ spin-exchange
anisotropy has a far more elaborate effect on its
low-energy physics than previously appreciated. Below
we briefly summarize our main findings and discuss
their implications. A detailed account of our
results is presented in Fig.~\ref{fig:fig1}.

We begin with a spin-$\frac{1}{2}$ impurity and with
$k > 2$ conduction-electron channels. From conformal
field theory it is known that the non-Fermi-liquid fixed
point of the corresponding Kondo model is stable against
a small spin-exchange anisotropy.~\cite{ALPC92} However,
it was found to be unstable against a sufficiently
large $J_z \gg |J_{\perp}| > 0$. In the latter regime,
the system flows to a line of stable ferromagnetic-like
fixed points with a residual isospin-$\frac{1}{2}$
local moment. The phase diagram of the model thus
consists of three distinct domains:
the conventional ferromagnetic and antiferromagnetic
(i.e., non-Fermi-liquid) domains, plus a second
ferromagnetic-like domain located deep in the
antiferromagnetic regime. The new ferromagnetic-like
domain extends above a critical longitudinal coupling
$J_z^{\ast}$, whose magnitude depends on $k$. While
$J_z^{\ast}$ exceeds the bandwidth for intermediate
values of $k$, it is pushed to weak coupling for
$k \gg 1$. Each of the two ferromagnetic-type domains
is separated from the antiferromagnetic one by a
Kosterliz-Thouless line, as depicted in
Fig.~\ref{fig:fig1}(a). While we cannot entirely
rule out the possibility of yet another domain for
sufficiently large $|J_{\perp}| \gg J_z$, there are
no indications at this point in favor of such a
scenario.

Proceeding to $1/2 < s < k/2 - 1/2$, spin-exchange
anisotropy is known~\cite{ALPC92} to be a relevant
perturbation at the overscreened non-Fermi-liquid
fixed point for $k > 4$ (for $k = 4$ it is a marginal
perturbation). The basin of attraction of the
overscreened fixed point is therefore confined to
the line $J_z = |J_{\perp}| > 0$. The nature of the
low-temperature fixed points for $J_z \neq |J_{\perp}|$
was never explored. As shown in
Secs.~\ref{sec:Perturbative-RG} and \ref{sec:SC-general-s},
the system flows to a line of stable ferromagnetic-like
fixed points with a residual isospin-$\frac{1}{2}$ local
moment both for sufficiently small $J_z > |J_{\perp}|$
(i.e., in the limit $J_z, J_{\perp} \to 0$ for any given
ratio $J_z/|J_{\perp}| > 1$) and for a sufficiently large
$J_z \gg |J_{\perp}| > 0$. This suggests a single generic
behavior in the entire domain $J_z > |J_{\perp}|$.
Further support in favor of this interpretation is
provided in Appendix~\ref{app:s=1}, for the special
case of a spin-one impurity. As for the domain
$|J_z| < |J_{\perp}|$, here information is confined to
the weak-coupling regime, $|\rho J_{\perp}| \ll 1/ks$.
Depending on the parity of $2s$, the system
flows either to a conventional Fermi liquid with no
residual degeneracy (integer $s$), or to a $k$-channel
Kondo effect with an effective spin-$\frac{1}{2}$ local
moment (half-integer $s$). It remains to be seen to what
extent is this behavior generic to $|J_z| < |J_{\perp}|$.

For isotropic antiferromagnetic exchange and $k > 2$, the
two overscreened spins $s = 1/2$ and $s = k/2 - 1/2$ share
the same low-energy physics. Indeed, the corresponding
Kondo models are related for $J_z = J_{\perp} > 0$ through
a weak-to-strong-coupling duality. It is not surprising,
then, that the overscreened fixed point shows the same
stability for both spins against a weak spin-exchange
anisotropy.~\cite{ALPC92} A similar duality appears to
hold also for an XXZ anisotropy, at least in the range
$J_z > |J_{\perp}|$. Indeed, for $s = 1/2$ the
overscreened fixed point is stable at weak coupling
(assuming $J_z > -|J_{\perp}|$), giving way to a
line of stable ferromagnetic-like fixed points for
sufficiently large $J_z \gg |J_{\perp}| > 0$. For
$s = k/2 - 1/2$ the roles are reversed. The overscreened
fixed point is stable against a sufficiently large
$J_z \gg |J_{\perp}| > 0$, but is unstable for
sufficiently small $J_z > |J_{\perp}|$. In the latter
regime, the system flows to the same line of stable
ferromagnetic-like fixed points that is approached
for $s = 1/2$ and a large $J_z$. Note that, for
$s = k/2 - 1/2$ and small $|J_{\perp}| > |J_z|$, the
stability of the overscreened fixed point depends on
the parity of $k$. The overscreened fixed point is
stable for half-integer $s$ (even $k$), but is unstable
for integer $s$ (odd $k$).

All cases discussed above pertain to an overscreened
impurity. We now turn to an underscreened spin, i.e.,
$s > k/2$. As shown in Secs.~\ref{sec:Perturbative-RG}
and \ref{sec:SC-general-s}, an XXZ anisotropy is a
relevant perturbation both near the free-impurity
fixed point and for a sufficiently large
$J_z \gg |J_{\perp}|$. The sole exception to the
rule is the case $s = k/2 + 1/2$, where
$\delta J = J_z - |J_{\perp}| > 0$ is a marginal
perturbation in each of these limits. Consider first
the range $J_z > |J_{\perp}|$. Here the flow in both
extremes is to the same line of ferromagnetic-like
fixed points with a residual isospin-$\frac{1}{2}$
local moment. As before, this suggests a single generic
behavior throughout the domain $J_z > |J_{\perp}|$.
Similar to the case of isotropic antiferromagnetic
exchange, the low-energy physics is comprised of
quasiparticle excitations plus a residual local moment.
However, the residual degeneracy for $s > k/2 + 1/2$
is smaller than that of the isotropic underscreened
fixed point (two versus $2s - k +1 > 2$), distinguishing
the ferromagnetic-like line of fixed points from the
isotropic underscreened one.

Of the different possible cases, the most intriguing
perhaps is that of an underscreened impurity with
$|J_z| < |J_{\perp}|$ and half-integer $s$. As we
have shown in Sec.~\ref{sec:Perturbative-RG}, the
resulting low-energy physics is that of a $k$-channel,
spin-$\frac{1}{2}$ Kondo effect, at least in the limit
of sufficiently weak coupling. Thus, an XXZ anisotropy
drives the system from underscreened to overscreened
behavior. A more complete characterization of the
transition between these two distinctly different
behaviors is clearly needed.

Going back to $s = 1/2$, we conclude with a few further
comments on the mapping of Eq.~(\ref{mapping}). We
first reiterate that $J_z^{\ast}$ lies well beyond
the strict range of validity of bosonization for
intermediate values of $k$. Nevertheless, this approach
(and its Anderson-Yuval equivalent~\cite{Ye96}) well
describes the new ferromagnetic-like phase both
qualitatively and quantitatively. For $k = 3$, which
features the largest $J_z^{\ast}$ and is thus the most
prone to error, Eq.~(\ref{J-ast}) is only a factor of two
larger than the NRG estimate for a symmetric box density
of states. We find this degree of quantitative agreement
to be quite remarkable. 

Although the mapping of Eq.~(\ref{mapping}) was derived
in Sec.~\ref{sec:mapping} for a channel-isotropic model,
it can easily be extended to channel anisotropy both in
the spin-flip coupling, $J_{\perp} \to J_{\perp n}$, and
in the longitudinal coupling, $J_{z} \to J_{z n}$. The
individual transverse couplings remain unchanged in the
course of the mapping, whether isotropic or not. As for
the individual longitudinal couplings, these transform
according to
\begin{equation}
\arctan \left (\frac{\pi \rho J'_{z n}}{4} \right ) =
      \frac{\pi}{k}
      + \arctan \left (\frac{\pi \rho J_{z n}}{4} \right )
      - 2 \bar{\delta}_z ,
\label{anisotropic_mapping-1}
\end{equation}
where
\begin{equation}
\bar{\delta}_z = \frac{1}{k} \sum_{n = 1}^{k}
      \arctan \left (\frac{\pi \rho J_{z n}}{4} \right )
\label{anisotropic_mapping-2}
\end{equation}
is the average phase shift for the $k$ different
channels. Accordingly, the mapping of
Eqs.~(\ref{anisotropic_mapping-1}) and
(\ref{anisotropic_mapping-2}) is restricted to values
of $J_{z n}$ where the modulus of the right-hand side of
Eq.~(\ref{anisotropic_mapping-1}) does not exceed $\pi/2$
for any of the $k$ channels. Observe that channel anisotropy
is preserved by Eqs.~(\ref{anisotropic_mapping-1}) and
(\ref{anisotropic_mapping-2}), which adequately reduce to
Eq.~(\ref{mapping}) in the limit of isotropic couplings.

Finally, we remark on the possibility of generalizing
the mapping of Eq.~(\ref{mapping}) to arbitrary spin $s$.
Two modifications appear when the same sequence of steps
is applied to an impurity spin larger than one-half:
(i) For $s > 1$, the phase shift $\delta_z$ in the absence
    of $J_{\perp}$ is a nonlinear function of $S_z$. Hence,
    the bosonized form of the $J_z$ interaction term is no
    longer linear in $S_z$ as for $s = 1/2$.
(ii) The unitary transformation $U$ produces an additional
     Hamiltonian term of the form $\Delta S_z^2$, similar
     to the one generated in perturbative RG [see
     Eq.~(\ref{d-Delta-dl})]. For $s = 1/2$, this term
     amounts to a uniform shift of the entire spectrum,
     which can be safely ignored.  This, however, is no
     longer the case for $s > 1/2$, where different
     Kramers doublets are split.
As a result of the former modification, the mapped Hamiltonian
no longer assumes the form of a simple spin-exchange
Hamiltonian for $s > 1$. The case $s = 1$ is an exception
in this regard. The mapped Hamiltonian does acquire an
additional $\Delta S_z^2$ term, but otherwise retains
the form of a conventional spin-exchange interaction. A
detailed discussion of this particular case is presented
in Appendix~\ref{app:s=1}.

\section*{Acknowledgments}

Stimulating discussions with Natan Andrei, Piers Coleman,
Andres Jerez, Eran Lebanon, Pankaj Mehta, and Gergely
Zar\'and are gratefully acknowledged. A.S was supported
in part by the Centers of Excellence Program of the Israel
science foundation, founded by The Israel Academy of
Science and Humanities.

\appendix

\section{Exact mapping for $s = 1$}
\label{app:s=1}

In this appendix, we extend the mapping of Eq.~(\ref{mapping})
to a spin-one impurity. As explained in the main text, $s = 1$
is the only other spin size for which a simple spin-exchange
interaction is restored at the conclusion of the mapping.
However, an additional local field proportional to $S_z^2$
is generated. The basic steps of the derivation are nearly
identical to those carried out in Sec.~\ref{sec:mapping}
for $s = 1/2$. Only a few minor modifications appear, as
specified below.

Our starting point is the Hamiltonian of Eq.~(\ref{H}),
where $\vec{S}$ now represents a spin-one operator.
Bosonizing the fermion fields according to
Eq.~(\ref{psi-via-phi}), the bosonic Hamiltonian
assumes the form of Eq.~(\ref{H-bosonized}) with
one sole variation: The longitudinal spin-exchange
term now reads
\begin{equation}
\delta_1 \frac{a}{2 \pi^2 \rho} \sum_{n = 1}^{k}
             \left [
                    \nabla \Phi_{n \uparrow}(0)
                    - \nabla \Phi_{n \downarrow}(0)
             \right ] S_z ,
\end{equation}
where
\begin{equation}
\delta_1 = \arctan \left (\frac{\pi \rho J_z}{2} \right ) .
\end{equation}
Converting to the boson fields of Eqs.~(\ref{Phi_s-def})
and (\ref{Phi_mu-def}) and applying the transformation
$U = \exp \left [i \sqrt{\frac{8}{k}} \Phi_s(0) S_z \right]$,
the transformed Hamiltonian
${\cal H}' = U {\cal H} U^{\dagger}$ retains the same
overall form as Eq.~(\ref{H'}), but with two important
modifications:
(i) The coefficient of the $\nabla \Phi_s(0) S_z$
term is replaced with
\begin{equation}
\left( \delta_1 - \frac{2\pi}{k} \right)
              \frac{a}{2 \pi^2 \rho} \sqrt{2k} \; ;
\end{equation}
(ii) A new Hamiltonian term
${\cal H}_{\Delta} = \Delta S_z^2$ with
\begin{equation}
\Delta = 8D \left [
                    \frac{1}{k} - \frac{\delta_1}{\pi}
            \right]
\end{equation}
is added to ${\cal H}'$.~\cite{comment_on_delta} Proceeding
with the transformation $\Phi_s(x) \to -\Phi_s(x)$ and
converting back to a fermionic representation, the
Hamiltonian ${\cal H}'$ regains the form of Eq.~(\ref{H}),
but with certain renormalized parameters:
\begin{equation}
J_z \to J'_z ,
\label{mapping(a)-S=1}
\end{equation}
\begin{equation}
h \to -h ,
\end{equation}
\begin{equation}
g_i \to 2g_e - g_i ,
\end{equation}
and
\begin{equation}
\Delta = 0 \to 8D \left [ \frac{1}{k} -
                          \frac{\delta_1}{\pi}
                  \right ] .
\end{equation}
Here $J'_z$ is determined from the equation
\begin{equation}
\arctan \left (\frac{\pi \rho J'_z}{2} \right ) =
      \frac{2\pi}{k}
      - \arctan \left (\frac{\pi \rho J_z}{2} \right ) ,
\label{mapping(b)-S=1}
\end{equation}
which comes in place of Eq.~(\ref{mapping}) for a
spin-$\frac{1}{2}$ impurity.

\begin{figure}
\centerline{
\includegraphics[width=80mm]{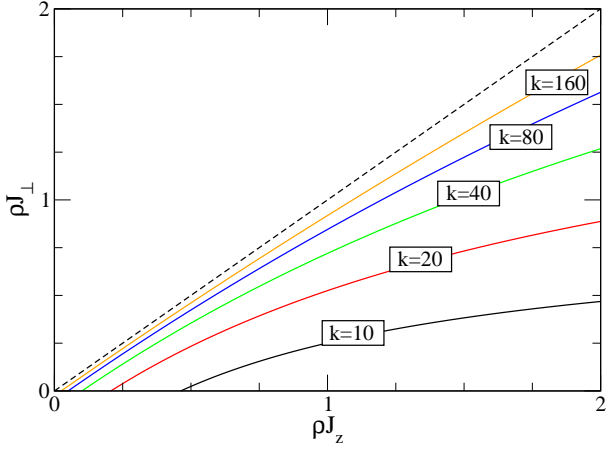}
}\vspace{-5pt}
\caption{(Color online)
         Lower bound on the basin of attraction of the line
         of stable ferromagnetic-like fixed points, for $s = 1$
         and different values of $k \gg 1$. In the region
         below and to the right of each solid line, both
         $\Delta < 0$ and $J'_z < - |J_{\perp}|$. Hence the
         system flows to the line of stable ferromagnetic-like
         fixed points with a residual degeneracy of two. The
         fixed-point structure in the remaining portion of
         the domain $J_z > |J_{\perp}|$, i.e., in between
         the solid lines and the overscreened line $J_z =
         |J_{\perp}|$ (marked by a dashed line), cannot
         be deduced based on
         Eqs.~(\ref{mapping(a)-S=1})--(\ref{mapping(b)-S=1})
         alone.}
\label{fig:fig5}
\end{figure}

As for $s = 1/2$, the mapping of
Eqs.~(\ref{mapping(a)-S=1})--(\ref{mapping(b)-S=1}) is
restricted to values of $J_z$ where the right-hand side
of Eq.~(\ref{mapping(b)-S=1}) does not exceed $\pi/2$.
This constrains the mapping to $k \ge 3$ (overscreened
impurity), and to $J_z \ge J_{\rm min}$ with
\begin{equation}
J_{\rm min} = \left \{
\begin{array}{lcc}
\frac{2}{\sqrt{3} \pi \rho} > 0 & \;\; & k = 3 \\ \\
0                               & \;\; & k = 4 \\ \\
-\frac{2}{\pi \rho}
\tan \left( \frac{\pi}{2} - \frac{2\pi}{k} \right ) < 0
                                & \;\; & k > 4
\end{array} \;\; .
\right.
\end{equation}
Specifically, for $k > 4$ the coupling regimes
$J_{\rm min} \leq J'_z < 0$ and
$J_z > J_z^{\ast} > 0$ with
\begin{equation}
J_z^{\ast} = \frac{2}{\pi \rho} \tan
                \left (
                      \frac{2 \pi}{k}
                \right )
\end{equation}
are mapped onto one another. This should be compared with
Eqs.~(\ref{J_min}) and (\ref{J-ast}) for $s = 1/2$.

Since Eqs.~(\ref{mapping(a)-S=1})--(\ref{mapping(b)-S=1})
define yet another multichannel Kondo Hamiltonian with
both spin-exchange anisotropy and a potentially competing
$\Delta S_z^2$ term, we have no conclusive way to deduce
its fixed-point structure throughout the $J_z$--$J_{\perp}$
plane. Nevertheless, the behavior in one particular
region is clear. For values of $J_z$ where both
$J_z' < - |J_{\perp}|$ and $\Delta < 0$,
the spin-exchange interaction and the
$\Delta S_z^2$ term conspire to favor a ferromagnetic-like
state with two-fold residual degeneracy. Here the residual
degeneracy originates from the $S_z^2 = 1$ Kramers doublet
favored by $\Delta < 0$. For such values of $J_z$, one can
safely conclude that the system flows to the line of stable
ferromagnetic-like fixed points identified previously from
the strong-coupling expansion of Sec.~\ref{sec:SC-general-s}.
These considerations provide us with the following lower
bound on the basin of attraction of the
ferromagnetic-like line of fixed points:
\begin{equation}
|J_{\perp}| < |J_{\rm min}| ,
\label{boundary-I}
\end{equation}
\begin{equation}
J_z > \frac{2}{\pi \rho} \tan
           \left [
                   \frac{2\pi}{k} +
                   \arctan \left(
                               \frac{\pi \rho |J_{\perp}|}{2}
                         \right )
           \right ] .
\label{boundary-II}
\end{equation}

Figure~\ref{fig:fig5} depicts that portion of the
$J_z$--$J_{\perp}$ plane defined by
Eqs.~(\ref{boundary-I}) and (\ref{boundary-II}), for
several values of $k \gg 1$. With increasing
$k$, a growing fraction of the domain $J_z > |J_{\perp}|$
is covered by this region, which stretches for $k \gg 1$
from $\rho J_z \approx 4/k \ll 1$ and upward in $J_z$.
This result partially bridges between the strong- and
weak-coupling limits ($\rho J_z \gg 1, |\rho J_{\perp}|$ and
$0 < |\rho J_{\perp}| < \rho J_z \ll 1/k$, respectively),
where flow to a ferromagnetic-like state has been established
in Secs.~\ref{sec:Perturbative-RG} and \ref{sec:SC-general-s}
using vastly different techniques. As for the remaining
fraction of the domain $J_z > |J_{\perp}|$, its fixed-point
structure cannot be immediately deduced based on
Eqs.~(\ref{mapping(a)-S=1})--(\ref{mapping(b)-S=1})
alone.


\begin{thebibliography}{99}
\bibitem{comment_on_trunctaion} Careless truncation may
     result in spurious symmetry breaking. To avoid symmetry
     breaking in the course of truncation it is essential to
     accommodate the degenerate multiplets of ${\cal H}_N$
     in their entirety. This may require a slight increase
     in the number of states retained at certain iterations.
\bibitem{comment_on_conserved_numbers} In our code we exploited
     the conservation of the following quantities: the total
     electronic charge, the $z$ component of the overall spin
     of the system, and the operator
     $L_z = \sum_{j = 1}^{N} \sum_{\sigma} \left [
         f^{\dagger}_{j 3 \sigma} f_{j 3 \sigma}
         - f^{\dagger}_{j 1 \sigma} f_{j 1 \sigma} \right ]$.
     We did not make use of the full SU(3) channel symmetry
     of the model.
\bibitem{comment_on_delta} $\Delta$ is formally proportional
     to $\delta(0)$, which is regularized as $1/a$.
\end{thebibliography}
\end{document}